
\documentstyle{amsppt}
\magnification\magstep1
\font\got=eufm10
\voffset 2 truecm
\vsize 8.5 truein

\overfullrule 0pt
\def\gotO{\text{\got A}}
\TagsOnRight
\frenchspacing
\baselineskip15pt

\topmatter
\rightline{DESY 94-186}
\rightline{hep-th/9410185}

\title Phase space properties of charged fields\\
in~theories~of~local~observables \\ \\
{\it D\lowercase{edicated to} B. S\lowercase{chroer}\/}\endtitle\bigskip
\author D. Buchholz\footnotemark"$^{a)}$" \  {\rm and} C.
D'Antoni\footnotemark"$^{b)}$"\footnotemark"$^{*}$"\endauthor
\affil
\footnotemark"$^{a)}$" {II. Institut f\"ur Theoretische Physik,
Universit\"at~Hamburg \ D--22761 Hamburg, Germany.}\\
\footnotemark"$^{b)}$"  {Dipartimento di Matematica, Universit\`a di
Roma \\``Tor Vergata'' \ I--00133 Rome, Italy.}
\endaffil

\bigskip
\abstract{Within the setting of algebraic quantum field
theory a relation between phase-space properties of
observables and charged fields is established. These
properties are expressed in terms of compactness and nuclearity
conditions which are the basis for the
characterization of theories with physically reasonable
causal and thermal features.
Relevant concepts and results of phase space analysis
in algebraic quantum field theory are reviewed and the
underlying ideas are outlined.} \endabstract
\endtopmatter

\footnotetext "$^{*}$"{Research supported by MURST.}

\leftheadtext{}
\rightheadtext{}

\vskip1cm
\document

\baselineskip13.5pt

\noindent{\bf 1. Introduction}

\noindent In the general structural analysis of relativistic quantum
field theories it has proved to be useful to characterize the phase
space properties of a theory with the help of compactness or
nuclearity conditions. These conditions have been a key to the
understanding of several physically significant issues, such as the
problem of causal (statistical) independence [1-5] and the existence
and structure of thermal equilibrium states [6-8]. They are also
an important ingredient in the analysis of the particle aspects of
the theory [9-12].

The heuristic basis of these conditions is the following physical
idea, which is almost as old as quantum theory itself [13]: In any
physically reasonable theory the number of states of limited total
energy and spatial extension should be finite because of the
uncertainty principle. Roughly speaking, it should be proportional
to the volume of phase space occupied by the states.

The mathematical formulation of this idea proved to be
difficult, however, since phase space is an ambiguous concept in
quantum field theory. A first and important step was taken by
Haag and Swieca who, starting from a tentative particle
interpretation of localized states, were led to the formulation of a
compactness criterion [9]. We state here this condition in an
equivalent but mathematically somewhat more convenient form.
For this purpose we consider for any given bounded spacetime
region $\Cal O$ and number $\beta>0$ the map $\Theta
^{(0)}_{\beta,\Cal O}$ from the $C^{\ast}$-algebra $\gotO
(\Cal O)$ of observables localized in $\Cal O$ [14]  into the
physical Hilbert space $\Cal H$, which is given by
$$
\Theta^{(0)}_{\beta,\Cal O} (A) = e ^{-\beta H}A\Omega\ , \quad
A \in \gotO (\Cal O)\ .\eqno (1.1)
$$
Here $H$ denotes the (positive) Hamiltonian and $\Omega \in \Cal
H$ the vacuum vector. Each map
$\Theta^{(0)}_{\beta,\Cal O}$ is linear and bounded (as a map
between the Banach spaces $\gotO (\Cal O)$ and $\Cal H$). The
criterion of Haag and Swieca then amounts to saying that in
theories with a reasonable particle interpretation these maps
should be {\it compact\/}. (Cf. the subsequent section for basic
concepts and results pertaining to such maps of ``almost finite
rank''.)

Haag and Swieca proved that their criterion is satisfied in massive
free field theory, but fails to hold in some generalized free field
theories which do not have a particle interpretation. In the meantime
their
criterion has also been established in massless free field theories
[15] and in some interacting theories in two dimensions [3, Sec. 4].
It provides a characterization of theories with
not too many local degrees of freedom.

In spite of its physical significance, the compactness criterion did
not prove to be the desired tool for a detailed structural analysis of
quantum field
theories, the main reason being that the criterion is only of a
qualitative nature.
Haag and Swieca proposed also a quantitative version of their criterion,
but their
bounds on the approximate dimension of the pertinent maps were too
conservative since they did not take into account the constraints
coming from particle statistics. It took almost
twenty years until the relevance of the latter point was
recognized.

Starting from another heuristical input based on
thermodynamical considerations, Buchholz and Wichmann [1]
came to the conclusion that in theories admitting thermal
equilibrium states the maps $\Theta^{(0)}_{\beta,\Cal O}$ should
not only be compact but also nuclear. Moreover, they argued that
the nuclear norm of these maps should, for small $\beta$ and
large $\Cal O$,  coincide with the partition function of the canonical
ensemble at temperature $\beta^{-1}$ in a container of size
corresponding to $\Cal O$. They also showed that their
nuclearity criterion is satisfied in free field theory, cf. also [15]. It
soon became clear that the properties of the maps
$\Theta^{(0)}_{\beta,\Cal O}$ can also be expressed in terms of
other mathematical concepts, such as the notions of $\varepsilon$-content,
approximate dimension, $p$-nuclearity etc. [3]. The
corresponding reformulations of
the nuclearity criterion are frequently more
convenient in applications. Although
they are not completely equivalent, they impose essentially the same
physical constraints on the underlying theories.

Another, in a sense dual criterion testing the phase space
properties of relativistic quantum field theories has been
proposed by Fredenhagen and Hertel [16], cf. also [17]. Instead of
considering local excitations of the vacuum whose energy is then
cut off as in relation (1.1), these authors proceed from states
(positive linear and normalized functionals on the algebra of all
observables) with limited energy. They argue that, in analogy to
the Haag--Swieca compactness criterion, the restrictions of these
states to any given local algebra $\gotO (\Cal O)$ should form a
compact subset of the dual space of $\gotO (\Cal O)$ in all theories
of physical interest. It was later shown by Buchholz and Porrmann
that also in this criterion the condition of compactness may be
strengthened to nuclearity [17]. These authors also  clarified the
relation between the various existing formulations of compactness
and nuclearity conditions. Roughly speaking, the conditions of
Fredenhagen--Hertel type impose somewhat stronger constraints
on the phase space properties of a theory than their counterparts
in terms of the maps $\Theta^{(0)}_{\beta,\Cal O}$. Cf. [17] for
precise statements.

The compactness and nuclearity criterions are normally stated
in terms of the local observables of a theory. But they can likewise
be posed for the local charge--carrying fields. One then
imposes compactness or nuclearity conditions on the maps
$\Theta_{\beta,\Cal O}$ which map the $C^{\ast}$-algebra $\Cal
F(\Cal O)$ of fields localized in $\Cal O$ into the physical Hilbert
space $\Cal H$ according to relation (1.1), where $\gotO (\Cal O)$
then has to be replaced by
the algebra $\Cal F (\Cal O)$. Since the local field
algebras contain the observable algebras as subalgebras, the maps
$\Theta^{(0)}_{\beta,\Cal O}$ can be recovered from
$\Theta_{\beta,\Cal O}$ by restriction,
$\Theta^{(0)}_{\beta,\Cal O} = \Theta_{\beta,\Cal O}\upharpoonright
\gotO (\Cal O)$. It follows that compactness or nuclearity
conditions on the maps $\Theta_{\beta,\Cal O}$ imply that
the maps $\Theta^{(0)}_{\beta,\Cal O}$ also have the respective
properties.

It is the aim of the present article to study the
opposite problem, namely the question of whether one can deduce
from compactness or nuclearity properties of the maps
$\Theta^{(0)}_{\beta,\Cal O}$ corresponding properties of the
maps $\Theta_{\beta,\Cal O}$. This question is, on one hand, of
interest for a deeper understanding of the relation between the
canonical and grand canonical ensembles. According to the
thermodynamical considerations in [1], the maps
$\Theta^{(0)}_{\beta,\Cal O}$ should provide information on the
properties of the canonical ensemble of states carrying the charge
quantum numbers of the vacuum, whereas
$\Theta_{\beta,\Cal O}$ is linked to the grand canonical ensemble
of states of arbitrary charge and zero chemical potential. As was
already mentioned, the nuclear norms of these maps are a
substitute for the partition functions of the respective ensembles.

A clarification of the relation between the compactness or
nuclearity properties of these maps would, on the other hand,
allow to deduce algebraic properties of charged (unobservable)
fields, such as the split property [18], from phase space properties
of the observables. It would thereby lead to a deeper
understanding of the physical significance of these algebraic
structures which seem to be of importance, e.g., in connection
with the formulation of a quantum Noether theorem [19,
20].\medskip

In the present investigation we start from two general
assumptions on the theory which can be expressed purely in
terms of observables. The first assumption concerns the
superselection structure: let $\Sigma$ be some index set labelling
the superselection sectors describing localizable charges [21], let
$d_\sigma$ be the statistical dimension of the sector $\sigma \in
\Sigma$ and let $m_{\sigma}$ be the lowest mass value in this
sector (cf. Sec. 3 for precise definitions). Then we assume that
$$
\sum_{\sigma \in \Sigma} d_{\sigma} e^{-\lambda m_{\sigma}}<
\infty \quad \text{ for all } \ \lambda > 0\ .\eqno (1.2)
$$
This condition has clearly to be satisfied if the grand canonical
ensemble with zero chemical potential is to exist for arbitrary
positive temperatures.

Our second assumption concerns the phase space properties of
bilocal excitations of the vacuum which are induced by
observables. To this end we consider the appropriately defined
algebra $\gotO (\Cal O_1\cup\Cal O_2)$ of bilocal operations in
the spacelike separated regions $\Cal O_1,\,\Cal O_2$ and the
corresponding maps
$\Theta^{(0)}_{\beta,\Cal O_1\cup\Cal O_2}$. We assume that
the compactness or nuclearity properties of these maps are not
affected if one keeps $\Cal O_1$, say, fixed and shifts $\Cal O_2$
to spacelike infinity (cf. Sec. 3 for the precise formulation of this
condition).

The physical significance of this condition can be
understood by appealing again
to the thermodynamical interpretation of the pertinent maps.
According to that
interpretation the map $\Theta^{(0)}_{\beta,\Cal O_1\cup\Cal O_2}$
is related to
the canonical ensemble, confined in two containers of
size proportional to $\Cal O_1$, respectively $\Cal O_2$,
which are connected by a thin tube. The tube has
the effect that, although the total charge of the ensemble is fixed,
the charges in
the individual containers can fluctuate. This is the familiar situation
discussed in
the passage from the canonical to the grand canonical ensemble. Our second
assumption thus expresses the idea that a separation of the containers
should
have no significant effects, i.e., the bilocalized ensemble
should stabilize. Again,
this  condition seems to be necessary if the grand canonical
ensemble is to exist.

Starting from these assumptions we are able to establish qualitative and
quantitative compactness properties of the maps
$\Theta_{\beta,\Cal O}$ from corresponding properties of the maps
$\Theta^{(0)}_{\beta,\Cal O_1 \cup \Cal O_2}$. In particular we
will show that if
the maps $\Theta^{(0)}_{\beta,\Cal O_1 \cup \Cal O_2}$ are
of type $s$ (cf. Sec. 2),
then the maps $\Theta_{\beta,\Cal O}$ have this property, too, and
consequently the field algebra has the split property [3]. Thus the
present results reveal an intimate relation between the phase space
properties of charged fields and the underlying
observables. It is an interesting question whether the additional
assumptions on the observables, mentioned
above, are necessary to establish these results. Yet this
problem is not touched upon in the present investigation.

Our paper is organized as follows. In Sec. 2 we recall basic
concepts from the theory of compact linear maps between Banach
spaces and establish some useful technical results. Section 3
contains a discussion of the general setting of algebraic quantum
field theory as well as of the more specific assumptions entering
into our analysis. There we also obtain a preliminary result on
some relevant map, following from the cluster theorem, which
will allow us to establish our main results in Sec. 4.
Some physically significant applications are outlined in the
concluding Section 5.\bigskip

\noindent{\bf 2. Compact maps}

\noindent In the present investigation we have to rely on various
concepts
and results from the theory of compact linear maps between
Banach spaces. Standard references on this subject are the books
[22, 23]. We recall here the basic definitions and collect some
useful results. The expert reader may skip this section and return
to it later for some technical details.

We begin by explaining our notation. Let $\Cal E$ be any (real or
complex) Banach space with norm $\|\cdot\|_{\Cal E}$. The
unit ball of $\Cal E$ is denoted by $\Cal E_1$ and the space of
continuous linear functionals on $\Cal E$ by $\Cal E^{\ast}$. Given
another Banach space $\Cal F$, we denote the space of continuous
linear maps $L$ from $\Cal E$ to $\Cal F$ by $\Cal L (\Cal E,\Cal F)$.
The latter space is again a Banach space with norm given by
$$
\| L \| \doteq \sup\{ \| L (E)\| _{\Cal F}:E
\in \Cal E_1\}\ .\eqno (2.1)
$$
A map $L \in \Cal L (\Cal E, \Cal F)$ is said to be compact if the
image of $\Cal E_1$ under the action of $L$ is a precompact subset
of $\Cal F$. A convenient measure which provides more detailed
information on the size of the range of compact maps is the notion
of $\varepsilon$-content.\medskip

\noindent{\it Definition\/}: Let $L\in\Cal L(\Cal E, \Cal F)$ be a
compact map and let, for given $\varepsilon > 0,\,N_L(\varepsilon)$ be
the maximal number of elements $E_i\in\Cal E_1,\, i=1,\,\cdots
N_L(\varepsilon)$ such that $\| L (E_i-E_k)\|>\varepsilon$ if
$i\not= k$. The number $N_L(\varepsilon)$ is called the
$\varepsilon$-{\it content\/} of $L$. (Note that $N_L(\varepsilon)$
is finite for all $\varepsilon > 0$ iff $L$ is compact [22].)
\medskip
It is clear that the $\varepsilon$-content of $L$ increases if
$\varepsilon$
decreases, and it tends to infinity if $\varepsilon$ approaches 0
(unless $L$ is the zero map).
If the rank of $L$ is equal to $n\,\in\,\Bbb N$  (i.e., if the range of
$L$ is an  $n$-dimensional subspace of $\Cal F$), then the
$\varepsilon$-content $N_L(\varepsilon)$ behaves for small $\varepsilon$
like
$\varepsilon^{-n}$. The maps $L$ which will appear in our investigation
have
$\varepsilon$-contents which behave, for small $\varepsilon$, typically
like
$e^{(M/\varepsilon)^q}$, where $M$ and $q$ are positive numbers. This
fact suggests to introduce the quantities (provided they exist)
$$
q_L = \limsup _{\varepsilon \searrow 0}{ln\, ln\, N_L(\varepsilon)\over
 ln\, 1/\varepsilon} \ ,\eqno (2.2)
$$
called the {\it order\/} of $L$, and
$$
M_L(q)=\sup_{\varepsilon >0} \varepsilon  \bigl(ln\
N_L(\varepsilon)\bigr)^{1/q}\ ,\ q>0\ .\eqno (2.3)
$$
It is noteworthy that $q_L$ is just the infimum of all $q$ for which
$M_L(q)$ is finite.

Another way of describing the detailed properties of a compact
map is based on the idea to check how well the image of the unit
ball ${\Cal E}_1$ under the action of the map fits into suitable finite
dimensional
subspaces $\Cal S \subset \Cal F$.\medskip

\noindent {\it Definition\/}: Let $L\in \Cal L (\Cal E, \Cal F)$ be a
compact map and let $n\in \Bbb N_{\, 0}$. The number
$$
\delta_L(n) \doteq \inf \bigl\{ \delta>0:L(\Cal E_1) \subset \Cal S +
\delta \cdot
\Cal F_1\,,\,\Cal S \subset \Cal F\,,\, \dim \Cal S \leq n\bigr\}
$$
is called the $n${\it -th diameter\/} (of the image of $\Cal E_1$
under the action) of $L$. In particular $\delta_L(0)=\Vert L \Vert$.
\medskip

A closely related idea is to measure how much a given map
deviates from a map of finite rank. The relevant concepts are
given in the following definition.\medskip

\noindent {\it Definition\/}: Let $L\in \Cal L (\Cal E,\Cal F)$.
\item{i)} For given $n\in \Bbb N_{\, 0}$, the number
$$
\alpha_L(n)\doteq \inf \bigl\{ \| L - L_n\| : L_n \in \Cal
L (\Cal E, \Cal F)\ ,\ \text{rank}\ L_n\leq n\bigr\}
$$
is called the $n${\it -th approximation number\/} of $L$.
In particular $\alpha_L (0) = \Vert L \Vert$.
\item{ii)} The map $L$ is said to be of {\it type $l^p$\/}, $p>0$, if
$$
[] L []_p =
\bigl(\mathop{\sum}\limits_n\alpha_L (n)^p\bigr)^{1/p}<\infty\ .
$$
\noindent
(As is indicated by the notation, the maps of type $l^p$ form, for
fixed $p$, a linear space, and $[]\cdot[]_p$ is a (quasi) norm
on this space, cf. [22].)
\medskip

It is apparent that $\delta_L(n)\leq\alpha_L(n)$, but
in general there does not hold
equality. As a matter of fact, it need not even be true that $\alpha_L(n)$
tends to $0$ for large $n$ if $\delta_L(n)$ does, i.e., if $L$ is a
compact map. The two quantities coincide, however, in the
important special case where $\Cal F$ is a Hilbert space.

\noindent\proclaim {Lemma 2.1} Let $\Cal F$ be  a Hilbert space and
let $L\in \Cal L(\Cal E,\Cal F)$. Then $\alpha_L (n) = \delta_L(n),\
n\in\Bbb N_{\, 0}$.\endproclaim
\noindent{\it Proof\/}: As already mentioned, there holds always
$\delta_L(n)\leq\alpha_L(n)$. The reverse inequality follows from
the fact that for any $n$-dimensional subspace $\Cal S$ of a
Hilbert space $\Cal F$ there exists an orthogonal projection $P_{\Cal
S}\in \Cal L(\Cal F,\Cal F)$ of rank $n$ which projects onto  $\Cal S$.
Hence
$$
\alpha_L(n)\leq\inf\bigl\{ \| L-P_{\Cal S} \cdot L\|: \Cal
S \subset \Cal F,\dim \Cal S = n\bigr\}= \delta_L(n)\ ,
$$
where the last equality follows from the fact that $\|
L(E)-P_{\Cal S} \cdot L(E)\|_{\Cal F}=\inf \bigl\{\|
L(E)-F\|_{\Cal F}:F\in \Cal S\bigr\}$. \hfill $\qed$

\medskip
We conclude our list of quantities which measure the properties of
compact maps with still another concept. In this variant one
studies the decompositions of a given map into the most
elementary ones, the maps of rank one, which are of the form
$e(\cdot)F$, where $e\in\Cal E^{\ast}$ and $F\in\Cal F$.\medskip

\noindent{\it Definition\/}: Let $L\in\Cal L(\Cal E, \Cal F)$ be a
map such that for suitable sequences $e_n\in\Cal E^{\ast}\ ,\
F_n\in\Cal F\ ,\ n\in\Bbb N$ there holds (in the sense of strong
convergence)
$$
L(E)=\mathop{\sum}\limits_ne_n(E)F_n\ ,\ E\in\Cal E\ .
$$
If in addition there holds $\mathop{\sum}\limits _n\|
e_n\|^p_{\Cal E^{\ast}}\| F_n\|^p_{\Cal F}<\infty$ for
some $p>0$ the map $L$ is said to be $p$-{\it nuclear\/}. The space
of $p$-nuclear maps is equipped with the (quasi) norm [22]
$$
\| L\|_p=\inf \bigl(\mathop{\sum}\limits_n\|
e_n\|^p_{\Cal E^{\ast}} \| F_n\|^p_{\Cal
F}\bigr)^{1/p}\ ,
$$
where the infimum is to be taken with respect to all possible
decompositions of $L$. We note that 1-nuclear maps are in
general called nuclear maps.
\medskip

The various notions mentioned above are related to each other,
although these relations are not very rigid. In the present
investigation the concept of $\varepsilon$-content will prove  to be
the
most useful one. Since the other notions are also frequently used
in the literature, we collect here some useful information about
their respective relations. Again we restrict attention to the case
where $\Cal F$
is a Hilbert space (cf. [3; Sec. 2] for the general case). The following
lemma is a slight improvement on the general results in [22; Sec.
9.6] for the special case at hand.

\noindent\proclaim{Lemma 2.2} Let $\Cal F$ be a Hilbert space, let
$L\in \Cal L(\Cal E,\Cal F)$ be a compact map, and let $\varepsilon>0$.
There hold for $m\in\Bbb N$ and sufficiently large $n\in\Bbb N$ such
that $\alpha_L(n)<\varepsilon/2$ the inequalities
$$
{2^m\alpha_L(0)\cdots\alpha_L(m-1)\over v_m m!\,\varepsilon^m}
\leq N_L (\varepsilon) \leq \left( {2\alpha_L(0)\over \varepsilon -
2\alpha_L(n)} + 1\right )^n
$$
where $v_m$ is the
volume of the unit ball in $\Bbb R^{\, m}$.\endproclaim\medskip

\noindent {\it Proof\/}: We discuss here only the case
where $\Cal F$ is a real Hilbert space, the argument for
complex Hilbert spaces is analogous. To verify the statement we
have to study the geometric properties of the set ${\Cal S}\doteq L(\Cal
E_1)\subset\Cal F$. Since $\Cal E_1$ is absolutely convex and $L$ is
linear
and compact, the set ${\Cal S}$ is also absolutely convex and
precompact.
In order to prove the first inequality we fix some $0<\delta<1$ and
pick vectors $\Phi_m\in {\Cal S}\ ,\ m\in \Bbb N_{\, 0}$, such that $\|
  (1-P_m)\Phi_m\|_{\Cal F}=(1-\delta)\alpha_L(m)$, where
$P_m$ is the orthogonal projection onto the subspace generated
by $\Phi_0,\dots \Phi_{m-1}$, and $P_0=0$. Because of absolute
convexity, ${\Cal S}$ contains the polygon
${\Cal S}_m=\left\{\mathop{\sum}\limits^{m-1}_{i=0}c_i\Phi i :
\mathop{\sum}\limits^{m-1}_{i=0} |c_i| \leq 1\right\}$. Let $N_m$ be
the
maximal number of vectors $\Psi_j\in {\Cal S}_m\,,\,j=1, \cdots N_m$
with mutual distances larger than $\varepsilon$. In order to get a lower
bound on $N_m$ we regard ${\Cal S}_m$ as a subset of $\Bbb R^{\, m}$ and
note that the collection of $m$-balls of radius $\varepsilon$ which are
centered at the points $\Psi_j\,,\,j=1, \cdots N_m$ provides a
covering of ${\Cal S}_m$. The volume of the polygon ${\Cal S}_m$ is
equal to
$2^m (1-\delta)^m\alpha_L (0)\cdots \alpha_L (m-1)/m!$. It is not
larger than $N_m\varepsilon^mv_m$, i.e., the volume of the $N_m$ balls
covering it. But $N_m\leq N_L (\varepsilon)$, hence the first inequality
in the statement follows since $\delta$ can be made arbitrarily
small.

To prove the second inequality we recall that
$\alpha_L(n)=\delta_L(n)$ since $\Cal F$ is a Hilbert space. Thus,
given $\delta >0$, there is some at most $n$-dimensional subspace
${\Cal S}_n\subseteq \Cal F$ and a corresponding orthogonal
projection $P_n$ such that $\| (1-P_n)\Phi\|_{\Cal F}
\leq (1+\delta)\alpha_L (n)$ for all $\Phi\in {\Cal S}$. Hence if there
                                                   are
$N$ vectors $\Phi_i \in {\Cal S}$, $ i=1, \cdots  N$, such that $\|
\Phi_i-\Phi_k\|_{\Cal F}>\varepsilon$ for $i\not= k$, there holds
$\| E_n \Phi_i - E_n \Phi_k\|_{\Cal F}>\varepsilon -
2(1+\delta)\alpha_L(n)=\varepsilon'$.
If $n$ is such that $\varepsilon'>0$ we conclude, since $\|
E_n\Phi\| _{\Cal F}\leq\|\Phi\|_{\Cal F}\leq
\alpha_L(0)$ for $\Phi\in {\Cal S}$,
that there are not less than $N$ elements
in
the set $E_n\cdot {\Cal S}\subset\alpha_L(0)\bigl({\Cal S}_n\bigr)_1$
                        with
mutual distance larger than $\varepsilon'$. Hence there are not less
than
$N$ disjoint $n$-balls of radius $\varepsilon'/2$ fitting into the
$n$-ball
$\bigl(\alpha_L(0)+\varepsilon'/2\bigr)\cdot\bigl(\Cal S_n\bigr)_1$.
Comparing volumes as in the preceding step and making $\delta$
arbitrarily small one arrives at the second
inequality. \hfill $\qed$\medskip

In the analysis of the compactness properties of maps it is often
most convenient to determine their nuclear (quasi) norms. The
problem is then to infer from the size of these norms on the
$\varepsilon$-content and vice versa. In order to establish such
a relation we
need the following two lemmas which are of interest in their own right.

\noindent\proclaim{Lemma 2.3} Let $L_1,\dots  L_n\in \Cal L (\Cal E
, \Cal F)$ be compact maps with $\varepsilon$-content
$N_{L_1}(\varepsilon),\mathbreak
\cdots N_{L_n}(\varepsilon)$, respectively. Then the $\varepsilon$-content
of the map $L_1+\cdots + L_n$ is bounded by
$$
N_{L_1+\cdots + L_n} (\varepsilon) \leq
\inf_{\varepsilon_1+\cdots+\varepsilon_n=
\varepsilon/2} N_{L_1}(\varepsilon_1)\cdots N_{L_n}(\varepsilon_n)\ .
$$
\endproclaim\medskip

\noindent{\it Proof\/}: According to the very definition of $\varepsilon
$-content there exist for given $\varepsilon_m>0\,,\,m=1,\cdots n$,
exactly $N_{L_m}(\varepsilon_m)$ elements $E_{m,k} \in \Cal E_1\,,\, k=1,
\cdots
N_{L_m}(\varepsilon_m)$, such that for any $E\in \Cal E_1$ there
holds
$$
\| L_m \bigl(E-E_{m,k_m(E)}\bigr) \|_{\Cal F}\leq\varepsilon_m
$$
for some suitable index $k_m(E)$. Let $\Bbb I $ be the set of all
$n$-tuples $\bigl(k_1(E),\cdots k_n(E)\bigr),\,
E\in \Cal E_1$, which appear in this way. The cardinality of
$\Bbb I$ is less than or equal to $N_{L_1} (\varepsilon_1)\cdots N_{L_n}
(\varepsilon_n)$. For each $(i_1,\cdots i_n)\in \Bbb I $ we pick some
operator $\bar E_{ i_1,\cdots i_n} \in \Cal E_1$ such that $\| L_m
\bigl(\bar E_{ i_1, \cdots i_n} - E_m, i_m\bigr)\|_{\Cal F}
\leq \varepsilon_ m$
for
$m=1,\dots n$. Then there holds for any $E\in\Cal E_1$
$$
\align
\| \bigl(L_1&+\cdots+L_n\bigr)\bigl(E-\bar E_{i_1,\cdots
i_n}\bigr) \|_{\Cal F} \leq \sum ^n_{m=1} \| L_m \bigl(
\bar E - E_{i_1,\cdots i_n}\bigr)\|_{\Cal F}\\
&\leq \sum ^n_{m=1} \| L_m \bigl(E-E_{m, i_m}\bigr)\|_{\Cal
F} + \sum ^n_{m=1} \| L_m \bigl(E_{m,i_m}-\bar E_{i_1,\cdots
i_n}\bigr)\|_{\Cal F}\ .
\endalign
$$
Consequently there exists some index $\bigl(i_1,\cdots i_n\bigr) \in
\Bbb I$ such that
$$
\|\bigl(L_1+\cdots+L_n\bigr) \bigl(E-\bar E_{i_1,\cdots
i_n}\bigr)\|_{\Cal F} \leq 2 \bigl(\varepsilon_1+\cdots
+\varepsilon_n\bigr)\ .
$$
Setting $\varepsilon_1+\cdots+\varepsilon_n=\varepsilon/2$ we
conclude that the
$\varepsilon$-content of the map $\bigl(L_1+\cdots+L_n\bigr)$ cannot
be
larger than the cardinality of $\Bbb I$ , so the statement
follows. \hfill $\qed$

\noindent\proclaim{Lemma 2.4} Let $L\in \Cal L(\Cal E, \Cal F)$
and let $L_n\in \Cal L\bigl(\Cal E, \Cal F_n\bigr),\ n\in \Bbb N$ , be a
sequence of maps which dominates $L$ asymptotically, i.e., the
sequences $\| L_n(E)\|_{\Cal F_n}, E\in\Cal E,$
converge and
$$
\lim_n\| L_n(E)\|_{\Cal F_n}\geq \|
L(E)\| _{\Cal F}\ ,\ E\in\Cal E\ .
$$
If all maps $L_n$ are compact and have uniformly bounded $\varepsilon
$-contents, then $L$ is also compact, and
$$
N_L(\varepsilon)\leq \liminf_n N_{L_n}(\varepsilon)\ ,\
\varepsilon > 0\ .
$$\endproclaim\medskip

\noindent{\it Proof\/}: Let $E_i\in\Cal E_i\,,\, i=1\,,\,\cdots N$ be
such that $\| L(E_i)-L(E_k)\|_{\Cal F}>\varepsilon$ for
$i\not=k$. Since the sequence $L_n,n\in\Bbb N$ dominates $L$
asymptotically there exists some $n_0$ such that for $n\geq n_0$
there holds
$\| L_n (E_i) - L_n (E_k)\|_{\Cal F_n} > \varepsilon$ if
$i\not=k$.  Hence $N\leq N_{L_n}(\varepsilon)$ for all $n\geq n_0$ and
consequently $N\leq \liminf\limits_n N_{L_n}(\varepsilon)$. \hfill $\qed$

In the following proposition, which is partly taken from [22], some
useful relations between the $\varepsilon$-content (respectively the
related quantity $M_L(q)$ introduced in (2.3)), the $l^p$-norms
and the nuclear norms of a map $L$ are established.

\noindent\proclaim{Proposition 2.5} Let $\Cal F$ be a Hilbert space
and let $L\in\Cal L (\Cal E,\Cal F)$. There hold the inequalities

{\parindent25pt
\item{\rm i)} $\| L\|_p\leq a_p\cdot []
L[]_p\quad\text {for}\quad 0<p\leq 1$\smallskip
\item{\rm ii)} $[] L[]_p\leq b_{p,q}\cdot M_L(q)\quad\text
{for}\quad p> {2q/(2-q)}>0$\smallskip
\item{\rm iii)} $M_L(q) \leq c_{p,q}\cdot{}\| L \|_p\quad\text
{for}\quad q>{p/(1-p)}>0$.

\noindent
Here $a_p,\,b_{p,q}$ and $c_{p,q}$ are numerical constants which
do not depend on $L$.\endproclaim\medskip

\noindent{\it Proof\/}: i) The first inequality has been established
in [22, Prop. 8.4.2]. As a matter of fact it holds for maps between
arbitrary Banach spaces. For the constant in the inequality one
has the upper bound $a_p\leq 2^{2+3/p}$.

ii) According to the definition of $M_L(q)$ there holds $N_L(\varepsilon)
\leq e^{(M_L(q)/\varepsilon)^q}$ and consequently
$$
\inf_{\varepsilon>0} \varepsilon^n N_L(\varepsilon) \leq \left({q^{1/q}
e^{1/q}M_L(q)\over n^{1/q}}\right)^n\ ,\ n\in\Bbb N\ .
$$
Making use of the first inequality in Lemma 2.2 and the fact that
the approximation numbers are monotonously decreasing,  we
conclude that
$$
\alpha_L(n-1)\leq{(v_n n!)^{1/n}\over 2}
{q^{1/q} e^{1/q}\over n^{1/q}}\cdot M_L(q),\ n\in\Bbb N\ .
$$
But $(v_n n!)^{1/n} \leq \sqrt{2\pi n}$, so the statement
follows after summation. Moreover, one obtains a bound on the
constant $b_{p,q}$ given by $b_{p,q} \leq \sqrt {\pi\over 2} q^{1/q}
e^{1/q}\cdot\left({p\over p-2q/(2-q)}\right) ^{1/p}$.

iii) For given $\delta>0$ there exists a sequence of maps
$L_n\in\Cal L(\Cal E,\Cal F),\,n\in\Bbb N$ , of rank one such that
$L(E)=\sum\limits_nL_n(E)\, , \, E\in\Cal E_1$, in the sense of
strong convergence, and $\bigl(\sum\limits_n\|
L_n\| ^p\bigr)^{1/p}\leq (1+\delta)\cdot \|
L\|_p $ . We assume in the following that the sequence of norms $\|
L_n\|$ is monotonically decreasing.  (This can always be
accomplished by reordering of the operators $L_n$ since the sum
in the decomposition of $L$ is absolutely converging for
$p$-nuclear operators, $0<p\leq 1$.) It then follows that
for any given $n\in \Bbb N$ there holds $n\cdot\|
L_n\|^p \leq \sum\limits^n_{m=1}\| L_m \| ^p
\leq (1+\delta)^p\| L\|_p^p$ and consequently
$$
\| L_n\| \leq {(1+\delta)\| L\|_p\over
n^{1/p}}\quad,\quad n\in \Bbb N\ .
$$
Since the maps $L_n$ are of rank 1 it is straightforward to
estimate their $\varepsilon$-contents $N_{L_n}(\varepsilon)$. In the real
linear case $N_{L_n}(\varepsilon)$ is not larger than max$(4\| L_n
\| /\varepsilon, 1)$, and in the  complex linear case
than max$(4^2\| L_n\|^2/\varepsilon^2 ,1)$. We
treat in the following the latter case, the former one is analogous.
Applying Lemma 2.3 to the map $(L_1+\cdots+L_n)$ we obtain the
bound
$$N_{L_1+\cdots+L_n}(\varepsilon) \leq  N_{L_1}(\varepsilon_1)\cdots
N_{L_n}(\varepsilon_n)\ ,
$$
provided $\varepsilon_1+\cdots+\varepsilon_n = {\varepsilon\over 2}$. This
constraint on $\varepsilon_1,\cdots\varepsilon_n$ can be relaxed to
$\varepsilon_1+\cdots+\varepsilon_n \leq {\varepsilon\over 2}$ since the
$\varepsilon $-contents increase if $\varepsilon$ decreases.

We fix $ r ,\ 1< r <1/p$, and set
$$
\varepsilon_i = \left ( { r \over  r -1} (1+\delta)^{p r }\|
L\|^{p r }_p\right)^{-1}\| L_i\|
^{p r }\cdot {\varepsilon\over 2}\ ,\ i=1, \cdots n\ .
$$
In view of the preceding bound on the norms $\|
L_i\|$ there holds $(\varepsilon_1+\cdots +\varepsilon_n)\leq
\varepsilon/2$.
Moreover, the numbers $4\| L_m \| /\varepsilon_m\ ,\
m=1 , \cdots n$, are monotonically decreasing. Let $n_{\ast}$ be
the largest index for which
$4\| L_{n_{\ast}}\|/\varepsilon_{n_{\ast}}>1$;
if there is no such index we put $n_{\ast}=n$. Since for $m>n_{\ast}$
there holds $N_{L_m}(\varepsilon_m)=1$ one gets
$$
\align
N_{L_1+}&_{\cdots +L_n}(\varepsilon) \leq\\
&\leq \varepsilon^{-2n_{\ast}}\cdot 4^{3n_{\ast}}\left( { r \over
 r -1} (1+\delta)^{p r }\|
L\|^{p r }_p\right)^{2n_{\ast}}\cdot
\prod^{n_{\ast}}_{m=1} \| L_m\|
^{2(1-p r )}\leq\\
&\leq \varepsilon^{-2n_{\ast}}\cdot (n_{\ast})!^{2 {p r -1\over p}}
\left (
8 { r \over  r  - 1} (1+\delta) \| L \|_p\right)^
{2n_{\ast}}.
\endalign$$
Taking the supremum of this upper bound with respect to
$n_{\ast}\in\Bbb N$ one finds after a straightforward calculation
that
$$
N_{L_1+\cdots+ L_n}(\varepsilon) \leq e ^{(c_{p,q}\cdot(1+\delta)\|
L \|_p/\varepsilon)^q}
$$
where $q=p/(1-p r )$ and $c_{p,q}\leq8(2e/q)^{1/q} {1\over
1-pq/(q-p)}$. Since the sequence of maps $(L_1+\cdots+L_n)\in
\Cal L (\Cal E, \Cal F)$, $n\in \Bbb N$ , converges strongly to $L$, it
dominates $L$ asymptotically in the sense of Lemma 2.4. Hence
the preceding bound holds also for $N_L(\varepsilon)$. The statement
then follows in the limit of arbitrarily small
$\delta$. (Note that for $1< r <1/p$ the corresponding parameter
$q$ runs through the given range.)\hfill $\qed$ \medskip

It follows from this result that the maps of order $0$ coincide with
the maps which are of type $l^p$ or $p$-nuclear for all $p>0$. Such
maps are said to be of {\it type\/} $s$.

We conclude this section with two results on precompact sets
which will be needed later. In these statements there appears
the notion of the
$\varepsilon$-content $N^{\Cal S}(\varepsilon)$ of a subset $\Cal S$
of a Banach
space $\Cal E$: it is the largest natural number of elements $S_i\in
\Cal S\ ,\ i=1,\cdots N^{\Cal S}(\varepsilon)$, such that $\|
S_i-S_k\| _{\Cal E}>\varepsilon$ for $i\not=k$.

\noindent\proclaim{Lemma 2.6} Let $\Cal E_n$ be an
$n$-dimensional complex Hilbert space, $n\in\Bbb N$ , and let $\Cal
B\subset\Cal L (\Cal E_n\, ,\, \Cal E_n)$ be a subset of operators of
norm not larger than $b>0$. Then the $\varepsilon$-content of $\Cal B$
is bounded by
$$
N^{\Cal B} (\varepsilon) \leq \left({2nb\over
\varepsilon}+1\right)^{n^2}\ .
$$
\endproclaim\medskip

\noindent{\it Proof\/}: Let $\Phi_i\in \Cal E_n\,,\,i=1,\cdots n$ be
an orthonormal basis. There holds for any operator $B \in \Cal L
(\Cal E_n\, ,\,\Cal E_n)$
$$
\| B\| \leq \bigl(\sum_{i,k}|(\Phi_i\,,\,
B\Phi_k)|^2\bigr)^{1/2}\ \leq\ n\| B\|\ .
$$
Hence, viewing the matrix elements $(\Phi_i,\,B\Phi_k),\
i,k=1,\cdots n$ as components of a vector in $ \Bbb C^{\, n^2}$ we
conclude that the $\varepsilon$-content of $\Cal B$ is not larger than the
$\varepsilon$-content of a ball of radius $n b$ in $ \Bbb C^{\, n^2}$.
                                                                     The
bound on $N^{\Cal B}(\varepsilon)$ then follows by comparing volumes
as in the proof of Lemma 2.2. \hfill $\qed$

\proclaim{Lemma 2.7} Let $\Cal E$ be a Banach space, let $\Cal I
\subset \Cal L(\Cal E,\Cal E)$ be a compact set of isometries which is
stable under taking inverses, and let $\Cal X$ be a subset of $\Cal E$
with the following property: for each $\varepsilon>0$ there exist at most
$N(\varepsilon)$ elements $X_i\in \Cal X\ ,\ i=1,\cdots N(\varepsilon)$,
such that
$$
\inf_{I\in\Cal I}\| X_i-I(X_k)\|_{\Cal E}>\varepsilon\ ,\
i\not=k\ .
$$
Then $\Cal X$ is precompact and its $\varepsilon$-content is
bounded by
$$
N^{\Cal X}(\varepsilon)\leq \inf_{\varepsilon_1+x \varepsilon_2=
{\varepsilon \over 2}} N(\varepsilon_1)N^{\Cal I}(\varepsilon_2)\ ,
$$
where $x=\sup\limits_{X\in\Cal X}\| X\|_{\Cal E}$
and $N^{\Cal I}(\varepsilon)$ is the $\varepsilon$-content of $\Cal
I$.\endproclaim\medskip

\noindent{\it Proof\/}:  By assumption there exist for given
$\varepsilon
_1>0$ \ $N(\varepsilon_1)$ elements $X_i\in\Cal X$, $i=1,\cdots
N(\varepsilon_1)$, such that for any $Y\in \Cal X$ there holds either
$\inf\limits_{I\in\Cal I}\| Y-I(X_i)\|_{\Cal E} \leq
\varepsilon_1$ or $\inf\limits_{I\in\Cal I} \|
X_i-I(Y)\|_{\Cal E} \leq \varepsilon_1$ for some suitable index $i$.
Since $\Cal I$ consists of isometries and is stable under taking
inverses, we may assume without loss of generality that the first
inequality holds. Moreover, since $\Cal I$ is compact, there exists
also some $I_Y\in \Cal I$ such that $\|Y
-I_Y(X_i)\|_{\Cal E}\leq\varepsilon_1$. It follows in particular that
$\| Y\|_{\Cal E} \leq \sup\limits_i
\| X_i\|_{\Cal E} + \varepsilon_1$, hence the set $\Cal
X$ is bounded, $x=\sup\limits_{X\in\Cal X}\| X\|_{\Cal
E}<\infty$.

Let $\varepsilon_2>0$. Then there exist $N^{\Cal I}(\varepsilon_2)$
elements
$I_k\ ,\ k=1,\cdots N^{\Cal I}(\varepsilon_2)$ such that for any
$I\in\Cal I$ there holds $\| I-I_k\| \leq\varepsilon_2$
for some suitable $k$. Hence we have for any $Y\in\Cal X$ the
estimate
$$
\| Y-I_k(X_i)\|_\Cal E \leq \| Y-I_{Y}
(X_i)\|_\Cal E + x \| I_{Y} - I_k\| \leq
\varepsilon_1 + x \varepsilon_2
$$
for a suitable index pair $i,k$. We proceed now as in the proof of
Lemma 2.4: let $\Bbb I$ be the set of indexes $(i,k)$ which appear
in this way if $Y$ runs through $\Cal X$. For each such index
we pick $Y_{ik}\in \Cal X$ such that $\|
Y_{ik}-I_k(X_i)\|_\varepsilon \leq \varepsilon_1+x\varepsilon_2$.
It follows that for any $Y\in \Cal X$ there exists some
$(i,k)\in\Bbb I$ such that $\| Y-Y_{ik}\|_\varepsilon \leq
2(\varepsilon_1+x\varepsilon_2)$. Setting
$2(\varepsilon_1+x\varepsilon_2)=\varepsilon$ and noting that the
cardinality of
$\Bbb I$ is at most $N(\varepsilon_1) N^{\Cal I}(\varepsilon_2)$
the statement follows. \hfill $\qed$\bigskip

\noindent{\bf 3. Observables, fields and compactness conditions}

\noindent We list in this section the standard assumptions made in the
theory of local observables [14] and formulate in precise terms
the more specific conditions indicated in the Introduction.
We then recall some
fundamental results on the superselection structure and the structure of
charge--carrying fields which have been derived from these assumptions by
Doplicher and Roberts [24]. Making use of these facts and the cluster
theorem we
will be able to establish some preliminary results on the compactness
properties of certain specific maps which are the basis for the
analysis in the subsequent section.

We suppose that the net $\gotO$ of local observables is concretely
given as an inclusion preserving mapping (hence the term net)
$$
\Cal O \longrightarrow \gotO (\Cal O), \quad \Cal O \in
\Cal K \eqno (3.1)
$$
from the set of all double cones $\Cal K$ in
Minkowski space $\Bbb R^{\, 4}$ to von Neumann algebras $\gotO (\Cal O)$
acting on the
(complex, separable)
vacuum Hilbert space $\Cal H_0$.

The net $\gotO$ is supposed to be local, i.e.,
$$
\gotO (\Cal O_1)\subset \gotO (\Cal O_2)'\quad\text{if}\quad \Cal
O_1\subset \Cal O_2'\ ,\eqno(3.2)
$$
where $\Cal O'$ denotes the spacelike complement of
$\Cal O$ and $\Cal B'$ the set of all bounded operators which
commute with a given set $\Cal B$ of bounded  operators on
the underlying Hilbert space.\medskip

We will also have occasion to consider algebras of observables for
space--time regions other than double cones. Given $\Cal O\in \Cal
K$ we define $\gotO (\Cal O')$ as the $C^{\ast}$-algebra which is
generated by all algebras $\gotO (\Cal O_1),\ \Cal O_1\subset\Cal O'$
and $\Cal O_1\in \Cal K$. Because of locality there holds $\gotO(\Cal
O) \subset \gotO (\Cal O')'$. But we make here the stronger
assumption (Haag duality)
$$
\gotO (\Cal O)=\gotO (\Cal O')'\quad,\quad\Cal O\in \Cal K\ .\eqno
(3.3)
$$

In the presence of spontaneously broken symmetries the net of
observables does not comply with this condition, cf. [25, 26]. One
should then identify $\gotO$ in the following with the so-called
dual net $\gotO^d\ ,\ \Cal O\longrightarrow \gotO^d(\Cal
O)\doteq \gotO(\Cal O')'$, which satisfies Haag duality under quite
general conditions [27].

We will also consider algebras corresponding to pairs of double
cones $\Cal O_1,\Cal O_2$. In the spirit of Haag duality we define
$\gotO (\Cal O_1 \cup \Cal O_2)$ as the largest algebra of operators
which commute with all observables in $\Cal O_1'\cap\Cal O_2'$,
$$
\gotO (\Cal O_1\cup\Cal O_2)=\bigl(\bigcup_{\Cal O\subset\Cal
O_1'\cap \Cal O_2'}\gotO (\Cal O)\bigr)'\ .\eqno (3.4)
$$
So, strictly speaking, the algebra $\gotO (\Cal O_1\cup\Cal O_2)$
should be regarded as an element of the dual net. But in order to
simplify the notation we omit the superscript $d$ since there is no
danger of confusion.

We assume that on the vacuum Hilbert space $\Cal H_0$ there
exists some continuous unitary representation $U_0$ of the
space--time translations $x=(\underline x,t)$ which acts
covariantly on the net $\gotO$,
$$
U_0(x)\  \gotO (\Cal O)U_0(x)^{-1}=\gotO(\Cal O+x)\ ,\eqno (3.5)
$$
has spectrum in the closed forward lightcone,
$sp\,U_0\subset\bar V_+$, and leaves the (up to a phase unique)
vacuum vector $\Omega\in \Cal H_0$ invariant. We also assume
that $\Omega$ has the Reeh--Schlieder property, i.e., the set of
vectors $\gotO(\Cal O)\Omega$ is dense in $\Cal H_0$ for each $\Cal
O \in\Cal K$.

Our first more specific assumption concerns the phase space
properties of the theory. It is expressed as follows.

\magnification\magstep1
\font\got=eufm10
\voffset.5truecm
\def\gotO{\text{\got A}}

\medskip
\noindent{\it Condition C\/}: Let $\beta > 0$, let $\Cal O_1,\Cal
O_2\in\Cal K$, and let $\Theta_{\beta,\Cal O_1\cup \Cal O_2}^{(0)}$
be the map from $\gotO (\Cal O_1\cup \Cal O_2)$ into $\Cal H_0$
given by
$$
\Theta^{(0)}_{\beta,\Cal O_1\cup \Cal O_2 }(A)\doteq e^{-\beta H_0}
A\Omega\ ,\ A\in \gotO (\Cal O_1\cup \Cal O_2)\ ,
$$
where $H_0$ is the (positive) generator of the time translations on
$\Cal H_0$. These maps are compact for any choice of $\Cal O_1,
\Cal O_2$. Moreover, if $N_{\Theta_{\beta,\Cal O_1\cup \Cal
O_2}}(\varepsilon)$ denotes the $\varepsilon$-content of
these maps there holds for any $\varepsilon > 0$
$$\liminf_{\underline x} N_{\Theta_{\beta,\Cal O_1\cup
(\Cal O_2+\underline x)}^{(0)}}  (\varepsilon) = \underline{N}_
{\beta,\Cal O_1, \Cal O_2}(\varepsilon) < \infty\ ,$$
where the limit is understood for translations $\underline x$
tending to spacelike infinity.
\medskip

This condition is a variant of the compactness criterion of Haag
and Swieca. We will later make more restrictive assumptions on
the upper bounds $\underline{N}_{\beta,\Cal O_1,\Cal
O_2}(\varepsilon)$ appearing in the statement.
If one assumes, for example, that the maps
$\Theta ^{(0)}_{\beta,\Cal O_1\cup(\Cal O_2 +\underline x)}$ are
$p$-nuclear
for fixed $0<p<1$ and that $\| \Theta ^{(0)}_{\beta,\Cal O_1\cup
(\Cal O_2 +\underline x)} \| _p \leq {const}$ for $\underline x$ tending
                                            to
spacelike infinity, it follows from the third statement in Proposition
2.5 that $\underline{N} _{\beta,\Cal O_1,\Cal O_2} (\varepsilon)\leq
e^{(M_q/\varepsilon)^q}$ for $q> {p\over 1-p}$ and some $M_q <
\infty$. Such nuclearity properties can easily be established in
free field theory, cf. the appendix in [3]. They are expected to hold
quite generally in theories of physical interest.

Our next assumption concerns the charged states of the theory.
We are interested here in states of finite energy carrying a
localizable charge. As has been discussed in [21], these states can
be characterized by a simple ``selection criterion'', testing their
localization properties. They are put together into superselection
sectors, i.e., equivalence classes of (pure) states inducing
equivalent representations of the observables. We label these
sectors by the elements of some index set $\Sigma$. There are two
physically significant data of the sectors which enter into our
second condition. First, each sector $\sigma \in \Sigma$ has an
intrinsically defined statistical dimension $d_{\sigma} \in \Bbb N$ ,
which specifies the order of (para-)statistics of the states in the
sector; for sectors with Bose or Fermi statistics one has
$d_{\sigma} = 1$ [21]. The second relevant quantity is the threshold
mass $m_{\sigma}$, i.e., the lower boundary of the spectrum of the
mass operator in the sector $\sigma$ (which may be positive or
0).\medskip

\noindent{\it Condition P\/}: For each $\lambda > 0$ there holds
$$
\Cal Z(\lambda) \doteq \sum_{\sigma} d_{\sigma} e ^{-\lambda
m_{\sigma}} < \infty
$$
where  the sum extends over all sectors $\sigma \in \Sigma$. (The
function $\Cal Z$ will be called {\it little partition function\/} as it
provides a lower bound for the grand partition function for zero
chemical potential. Actually it would be sufficient for our
analysis if $\Cal Z
(\lambda)<\infty$ for some sufficiently small $\lambda >0$.)\medskip

We emphasize that these postulates involve only the observables
of a theory, they do not depend on the existence of charge
carrying fields. But according to a deep result of Doplicher and
Roberts [24] such fields can always be {\it constructed\/} under
the above conditions. (The special assumptions $C$ and $P$ are not
needed there.) More precisely, there is a field net $\Cal F$,
$$
\Cal O \longrightarrow \Cal F (\Cal O)\ ,\ \Cal O \in \Cal K\tag
3.6
$$
of von Neumann algebras which act on a complex Hilbert space
$\Cal H$ containing $\Cal H_0$ as a subspace. The algebras $\Cal F
(\Cal O)$ are generated by an irreducible set of field operators
with normal Bose-- and Fermi commutation relations at spacelike
distances. On $\Cal H$ there is a continuous unitary
representation $U$ of the translations, extending $U_0$, which
satisfies the spectrum condition, $sp\,U\subset \bar V_+$, and acts
covarianly on the field net,
$$
U(x) \Cal F (\Cal O) U(x)^{-1} = \Cal F (\Cal O + x)\ . \tag 3.7
$$
Moreover, there is a compact group $G$ and a continuous, unitary
and faithful representation $V$ of $G$ on $\Cal H$ which
commutes with the translations $U$ and acts locally on the field
net,
$$
V(g) \Cal F (\Cal O) V (g)^{-1} = \Cal F (\Cal O)\ ,\ g\in G\ .\tag 3.8
$$
We call $G$ the (global) gauge group and the unitaries $V(g)$, $g\in
G$, gauge transformations.

The Hilbert space $\Cal H$ can be decomposed into subspaces $\Cal
H^{\sigma},\ \sigma \in \Sigma$, on which $V$ acts like a multiple of
an irreducible representation of $G$ of dimension $d_\sigma$
(i.e., the dimension of the respective representation of $G$
coincides with the statistical dimension of the sector $\sigma$). The
net of fixed points under
the gauge group, $\Cal O \longrightarrow \Cal F^G(\Cal O) = \Cal F
(\Cal O) \cap V(G)'$, leaves each $\Cal H^{\sigma}$ invariant and
its restriction to $\Cal H^{\sigma}$ coincides
(up to multiplicity) with the irreducible
representation of the net of observables $\Cal O \longrightarrow
\gotO (\Cal O)$ in the sector $\sigma$.
In particular,
the vacuum sector $\Cal H_0$ coincides with the set of
$G$-invariant vectors in $\Cal H$ and the restriction of the fixed
point net $\Cal F^G$ to $\Cal H_0$ coincides with the original net
$\gotO$ of observables on $\Cal H_0$. Thus the superselection
sectors $\sigma\in\Sigma$ are in one--to--one correspondence to the
irreducible representation of $G$ and $\Sigma$ may therefore be
identified with the spectrum (dual) $\hat G$ of $G$.

As discussed in the Introduction, we want to analyse the phase
space properties of the field net $\Cal F$ which, as explained, is
canonically associated with any local net $\gotO$ of observables. To
this end we have to study the compactness properties of the maps
$\Theta_{\beta,\Cal O}$ from $\Cal F(\Cal O)$ to $\Cal H$, given by
$$
\Theta _{\beta,\Cal O} (\psi) = e ^{-\beta H} \psi \Omega\ ,\ \psi \in
\Cal F (\Cal O), \tag 3.9
$$
where $H$ is the (positive) generator of time translations on $\Cal
H$. In order to carry out this analysis we have to make still
another assumption which amounts to the existence of a PCT
operator.\medskip

\noindent{\it Condition J\/}: There is an antiunitary involution $J$
on $\Cal H$ which commutes with the gauge transformations
$V(g),\ g\in G$, satisfies $JU(x)=U(-x) J$ as well as
$J\Omega=\Omega$ and acts geometrically on the net,
$$
J\Cal F(\Cal O) J= \Cal F(-\Cal O)\ ,\ \Cal O \in \Cal K\ .
$$
In contrast to standard quantum field theory, the existence of
such a $J$ may not follow from the very general assumptions
outlined above [28]. But it seems likely that it can be deduced from
a similar condition, involving only the observables, by a
combination of results in [29] on the action of the PCT--operator
on local morphisms of $\gotO$ with the methods developed in [30].

Let us turn now to the investigation of our problem. We begin by
introducing some further notations. Let $dg$ be the normalized
Haar measure on $G$ and let, for any bounded operator $B \in
\Cal B (\Cal H)$,
$$
M(B) = \int dg V(g)\, B\,V(g)^{-1}\ ,\tag 3.10
$$
the mean of $B$ with respect to the gauge group $G$. The
map $M\in \Cal L (\Cal B (\Cal H) , \Cal B (\Cal H))$
is obviously linear and has norm 1.\medskip

The elements of $\Cal F(\Cal O)\ ,\ \Cal O \in \Cal K$, will be denoted
by $\psi,\psi'$, and the space--time translated operators $\psi$ by
$\psi (x)\doteq U(x)\psi U(x)^{-1}$. We write $\psi
(\underline x)$ if the underlying translation is spacelike and $\psi
(t)$ if it is a time translation.

We also consider the direct products $\Cal F(\Cal O_1) \times \Cal
F(\Cal O_2)$, \ $ \Cal O_1,\Cal O_2 \in \Cal K$, and regard them as
normed vector spaces with norm given by
$$
\| \sum_{i,k} \psi^{(1)}_i\times \psi ^{(2)}_k \| = \inf \biggl\{
\sum_{l,m} \| \psi^{(1)'}_l \|  \| \psi ^{(2)'}_m \| : \sum_{l,m} \psi
^{(1)'}_l\times \psi ^{(2)'}_m = \sum _{i,k} \psi ^{(1)}_i \times \psi
^{(2)}_k \biggr\}\tag 3.11
$$
in an obvious notation. By completion we obtain Banach spaces,
but this is of no relevance in the sequel.

We fix in the following $\beta >0$ and some double cone $\Cal O \in
\Cal K$ centered at 0, so that $\Cal O=-\Cal O$, and consider a
family of maps $\Xi_{\underline x}\,,\,\underline x$ spacelike,
from $\Cal F(\Cal
O)\times \Cal F (\Cal O)$ into $\Cal H_0$. These maps are obtained
by composition of three linear maps which are defined as follows.
The first one maps $\Cal F(\Cal O) \times \Cal F(\Cal O)$ onto $\Cal
F(\Cal O) \times \Cal F (\Cal O+\underline x)$ and is fixed by $\psi
\times \psi' \longrightarrow \psi \times \psi'(\underline x)$. Since
the translations induce automorphisms of the field net this map is
an isomorphism. The second one maps $\Cal F(\Cal O) \times \Cal F
(\Cal O+\underline x)$ into $\gotO (\Cal O\cup (\Cal O+\underline
x))$ and is given by
$$
\psi\times\psi'(\underline x) \longrightarrow M(\psi \cdot\psi'
(\underline x)) \upharpoonright \Cal H_0\ .
$$
Since fields and observables  commute at spacelike distances and
since observables are invariant under the action of the mean $M$,
it follows that the range of this map is indeed contained in $\gotO
\bigl(\Cal O\cup (\Cal O + \underline x)\bigr)$. Moreover, there
holds $\|M ( \sum \psi_i \cdot \psi'_k(\underline x)) \| \leq \sum \|
\psi_i\|\| \psi'_k (\underline x)\|$, and since the left hand side of
this inequality does not depend on the particular decomposition of
the sum it is also clear that the map is bounded in norm by 1. The
third map is $\Theta^{(0)}_{\beta,\Cal O\cup (\Cal O + \underline
x)}$, defined in Condition C and assumed to be compact. By
composition of these maps we obtain the map
$\Xi_{\underline x}$\ ,
$$
\Xi_{\underline x} (\psi \times \psi') = e^{-\beta H_0} M \bigl(\psi
\cdot \psi' (\underline x) \bigr) \Omega\ . \tag 3.12
$$
Since the $\varepsilon$-content of a compact map does not
increase under composition with maps of norm less than or
equal to 1, the following result is an immediate consequence of the
preceding discussion.

\noindent\proclaim{Lemma 3.1} Let Condition C be satisfied and
let $\Xi_{\underline x},\ \underline x$ spacelike, be the family of
maps from $\Cal F (\Cal O)\times \Cal F (\Cal O)$ into $\Cal H_0$
defined above. Then each $\Xi_{\underline x}$ is compact and there
holds in the limit of large spacelike translations $\underline x$
$$
\liminf_{\underline x} N_{\Xi_{\underline x}} (\varepsilon)\leq
\underline{N} (\varepsilon)\ ,\ \varepsilon >0
$$
where $\underline{N} (\varepsilon) \doteq \underline{N}_{\beta,
\Cal O, \Cal O }{(\varepsilon)}$, cf. Condition
C.\endproclaim\medskip

In the next step we show that the preceding Lemma provides
relevant information on the map $\Xi_{\infty}$ from $\Cal F(\Cal
O)\times \Cal F (\Cal O)$ into $\Cal H \otimes \Cal H$, given by
$$
\Xi_{\infty} (\psi \times \psi') = \int dg \bigl(V(g)e^{- \beta H}
\psi \Omega\bigr) \otimes \bigl (V(g) e ^ {-\beta H} \psi'
\Omega\bigr) . \tag 3.13
$$
In the proof we make use of a weak form of the cluster theorem
which does {\it not\/} depend on the existence of a mass gap in the
theory.

\noindent\proclaim{Lemma 3.2} Let Condition C be satisfied and
let $\Xi_{\infty}$ be the map from $\Cal F (\Cal O)\times \Cal F (\Cal
O)$ into $\Cal H\times \Cal H$ introduced above. Then $\Xi_{\infty}$
is compact and
$$
N_{\Xi_{\infty}} (\varepsilon) \leq \underline{N} (\varepsilon)\ ,\
\varepsilon > 0
$$
with the same $\underline{N} (\varepsilon)$ as is the preceding
Lemma.\endproclaim\medskip

\noindent{\it Proof\/}: The result is based on the fact that the
maps $\Xi _{\underline x}$ dominate $\Xi _{\infty}$ asymptotically
in the sense of Lemma 2.4. To verify this statement let us consider
for arbitrary $\psi_1,\psi'_1,\psi_2,\psi'_2\in \Cal F(\Cal O)$ the
scalar product in $\Cal H $
$$\bigl( \Xi_{\underline x} (\psi_1\times
\psi'_1), \Xi_{\underline x} (\psi_2\times \psi'_2)\bigr) =
\int dg\bigl(\psi_1 \psi'_1 (\underline x) \Omega,\,
e^{-2\beta H}\ \psi^g_2 \psi{'}^g_2 (\underline x) \Omega \bigr)
\ .\tag 3.14 $$
 Here we introduced the notation $\psi^g=V(g) \psi V(g)^{-1}$ and
made use of the fact that (a) $dg$ is the Haar measure on $G$ and
therefore invariant under left and right action of the group, $(b)$
the gauge transformations $V(g)$ commute with translations and
(c) the vacuum vector $\Omega$ is invariant under the action of
$V(g)$. In order to calculate this scalar product in the limit of large
spacelike $\underline x$ we make use of the fact that $H$ is
non--negative. Thus, by the spectral theorem, we can write
$$
e^{-2\beta H}= {1\over \sqrt{2\pi}}\int dt f_{\beta} (t)
U(t)\ ,
$$
where $f_{\beta}$ is any test function whose Fourier transform
$\tilde f_{\beta}$ satisfies $\tilde f_\beta
(\omega) = e ^{-2\beta \omega}$ for $\omega \geq
0$. Hence the scalar product in (3.14) can be represented in the
form
$$
\int dg \int dt f_{\beta} (t) \bigl(\psi_1\psi'_1
(\underline x) \Omega\ ,\ \psi^g_2 (t) \psi{'}^g_2(t,\underline
x)\Omega \bigr)
$$
since $\Omega$ is invariant under translations. We now make use
of the spacelike (anti) commutation relations of the fields [24]:
every field
operator $\psi$ can be decomposed into a sum of operators of Bose and Fermi
type which commute, respectively anticommute at spacelike distances. Thus,
without restriction of generality, we may assume that the operators
$\psi_1$, $\psi'_1$, $\psi_2$, $\psi'_2$ are of fixed Bose or Fermi type.
We treat
here only the case where all operators are of Bose type,
the other cases are
analogous. Then, for fixed $t$ and sufficiently large $\underline x$,
there holds
$$
\bigl( \psi_1 \psi'_1 (\underline x) \Omega, \psi ^g_2 (t)
\psi{'}^g_2 (t,\underline x)\Omega\bigr)
= \bigl(\psi^g_2(t)^{\ast}\psi_1 \Omega\ ,\ U(\underline x)
\psi{'}^{\ast}_1\psi{'}^g_2(t) \Omega\bigr)\ ,
$$
where  we made again use of the fact that $\Omega$ is invariant
under translations. The latter expression converges in the limit of
large $\underline x$ to
$$
\bigl(\psi_1\Omega\ ,\ \psi^g_2 (t) \Omega \bigr)\bigl( \psi'_1
\Omega\ ,\ \psi{'}^g_2 (t) \Omega\bigr)
$$
since, as a consequence of locality, the spectrum of the generators
of spatial translations $U(\underline x)$ is Lebesgue absolutely
continuous, apart from a discrete part at $0$ corresponding to the
vacuum, cf. the arguments in [31, Sec. 2] which can easily be extended to
Fermi fields. Hence, applying the dominated convergence
theorem, we conclude that
$$
\align
\lim_{\underline x} &\ \bigl(\Xi_{\underline x} (\psi_1\times
\psi'_1)\,,\, \Xi_{\underline x} (\psi_2\times\psi'_2)\Omega
\bigr)=\\
&= \int dg \int dt\, f_{\beta} (t) \bigl( \psi_1
\Omega\, ,\,\psi^g_2 (t)\Omega\bigr) \bigl(\psi'_1\Omega \, ,\,
\psi{'}^g_2(t)\Omega\bigr)=\\ &= \int dg \bigl(\psi_1 \Omega \, ,\,
e^{-2\beta H_0}\psi^g_2\Omega\bigr)\bigl(\psi'_1 \Omega\, ,\,
e^{-2\beta H_0} \psi{'}^g_2 \Omega\bigr)\ , \endalign $$
where in the second equality we made again use of the positivity
of $H$ and the specific form of the function $f_{\beta}$. But
the last term in this equation coincides with
the scalar product $\bigl(
\Xi_{\infty}
(\psi_1 \times \psi'_1) \ ,\ \Xi_{\infty} (\psi_2\times \psi'_2)\bigr)$
in $\Cal H\otimes \Cal H$. Hence there holds
$$
\lim_{\underline x} \| \Xi_{\underline x} \bigl( \sum \psi_i \times
\psi'_k\bigr) \| = \| \Xi_{\infty} \bigl(\sum \psi_i \times
\psi'_k\bigr)\|
$$
for any $\sum \psi_i\times \psi'_k \in \Cal F (\Cal O) \times
\Cal   F(\Cal O)$. It follows that the (uniformly continuous) family
of maps $\Xi_{\underline x}$ dominates $\Xi_{\infty}$
asymptotically, as claimed. The statement now follows from Lemma
2.4 and the preceding lemma. \hfill $\qed$ \medskip

In a final preparatory step we proceed from $\Xi_{\infty}$ to a
map $\Xi$ from $\Cal F (\Cal O) \times \Cal F(\Cal O)$ to the space
$HS(\Cal H)$ of Hilbert Schmidt operators on $\Cal H$. We recall
that this space is equipped with the scalar product
$$
(H_1, H_2)_{HS}\doteq TrH^{\ast}_1H_2\ ,\quad H_1,H_2 \in HS
(\Cal H)\ .\tag 3.15
$$
The norm on $HS(\Cal H)$ will be denoted by $\|\quad\|_{HS}$. We
also recall that $HS(\Cal H)$ is canonically isomorphic to $\Cal
H\otimes \bar {\Cal H}$, where $\bar {\Cal H}$ is the conjugate
space of $\Cal H$. This isomorphism is established by assigning to
the elements $\Phi_1\otimes\bar {\Phi}_2\in \Cal H \otimes \bar
{\Cal H}$ the rank one operators $(\Phi_2,\cdot)\Phi_1\in HS (\Cal
H)$, and vice versa.\medskip

Let $\Xi$ be the linear map from $\Cal F (\Cal O) \times \Cal   F(\Cal
O)$ into $HS (\Cal H)$ given by
$$
\Xi (\psi\times\psi')=\int dg \bigl( V(g) e^{-\beta H}
\psi{'}^{\ast} \Omega,\cdot\bigr) V(g)e^{-\beta H}\psi \Omega\ .\tag
3.16
$$
The link between $\Xi$ and $\Xi_{\infty}$ is established with the
help of Condition J. Making use of the antiunitary involution $J$ and
the fact that $J \Cal H$ can be identified with $\bar {\Cal H}$,
we see that $\Xi$ is isomorphic to the map  from $\Cal F (\Cal O)
\times \Cal   F(\Cal O)$ into $\Cal H\otimes  \bar {\Cal H}$, given by
$$
\psi \times \psi' \to \int dg \bigl( V(g) e^{-\beta H}\psi
\Omega\bigr) \otimes \bigl(J\ V(g) e^{-\beta H}\psi{'}^{\ast} \Omega
\bigr)\ . \tag 3.17
$$
Since $J$ leaves $\Omega$ invariant and commutes with the gauge
group as well as all real functions of $H$, the right hand side of this
assignment can be rewritten in the form
$$
\int dg \bigl( V(g) e^{-\beta H} \psi \Omega\bigr) \otimes \bigl(
V(g) e^{-\beta H}J\psi{'}^{\ast}J\Omega\bigr)\ .\tag 3.18
$$
But with our choice of the region $\Cal O$ there holds $J\Cal F (\Cal
O)J = \Cal F(- \Cal O) = \Cal F(\Cal O)$, hence the map $\psi \times
\psi' \to \psi \times
J\psi{'}^{\ast} J$ defines an automorphism of the linear space $\Cal F
(\Cal O) \times \Cal   F(\Cal O)$. Thus we conclude that the maps
$\Xi$ and $\Xi_{\infty}$ are related by an isomorphism.
The following result is
then an immediate consequence.

\noindent\proclaim{Proposition 3.3} Let Conditions C and J be
satisfied and let $\Xi$ be the map from the space
$\Cal F (\Cal O) \times
\Cal   F(\Cal O)$ into $HS(\Cal H)$, defined above. Then $\Xi$ is
compact, and there holds
$$
N_{\Xi}(\varepsilon) \leq \underline{N} (\varepsilon)\ ,\
\varepsilon > 0\ ,
$$
with $\underline{N}(\varepsilon)$ as in Lemma
3.1.\endproclaim

\medskip
\medskip

\noindent{\bf 4. Compactness properties of field algebras}

\noindent The map $\Xi$, introduced in the preceding section, contains
the
desired information about the compactness properties of the map
$$
\Theta \ \doteq\ \Theta_{\beta,\Cal O} \tag 4.1
$$
as we shall demonstrate now. (Since $\Cal O$ and $\beta$ are kept fixed
in the following we can simplify the notation and omit these
subscripts.) Our strategy is as follows. We
decompose $\Cal H$ into the superselection sectors $ \Cal
H^{\sigma}_{\iota},\iota=1,\cdots d_{\sigma}$ (as already indicated,
each sector $\sigma$ appears with multiplicity $d_{\sigma}$) and
consider the corresponding maps $\Theta^{\sigma}_{\iota} =
P^{\sigma}_{\iota}\cdot \Theta$, where $P^{\sigma}_{\iota}$ is the
orthogonal projection onto $\Cal H^{\sigma}_{\iota}$. As we will see,
each of these maps is compact and has an $\varepsilon$--content
which is controlled by that of $\Xi$. In order to obtain information
on $\Theta$ we have to sum up the maps $\Theta^{\sigma}_{\iota}$,
and it is here where Condition P enters.  From that assumption it
follows that also the map $\Theta$ is compact and that its
$\varepsilon$--content is related to that of $\Xi$.

For any $\sigma \in \hat G$, corresponding to an irreducible
unitary representation of $G$ of dimension $d_{\sigma}$, we fix a
unitary matrix representation  $D^{\sigma}_{\iota \kappa} (\cdot)\, , \,
\iota\, , \, \kappa=1,\cdots d_{\sigma}$. With the help of this
representation we define the family of operators on $\Cal H$ given
by
$$
P^{\sigma}_{\iota} \doteq \int dg D^{\sigma}_{\iota\iota} (g) V(g)
\ .\tag 4.2
$$
It follows from the familiar orthogonality relations for unitary
matrix representations of compact groups [32] that the operators
$P^{\sigma}_{\iota}$ are orthogonal projections,
$P^{\sigma\ast}_{\iota} P^{\sigma '}_{\iota '} = \delta_{\sigma\sigma'}
\delta _{\iota \iota'} P^{\sigma}_{\iota}$. Moreover, each vector in
${\Cal H}^{\sigma}_{\iota} \doteq P^{\sigma}_{\iota} {\Cal H}$
transforms under
gauge transformations $V ( g )$ according to the irreducible
representation $\sigma$ of $G$. The orbits of these vectors span the
space ${\Cal H}^{\sigma} \doteq P^{\sigma} {\Cal H}$,
$P^\sigma \doteq \sum \limits^{d_{\sigma}}_{\iota=1} P^{\sigma}_{\iota}
$, i.e., the unique subspace of $\Cal H$ on which $V$ acts like a
multiple of the representation $\sigma$ and which is stable under
the action of the fixed point net ${\Cal F}^G$. Accordingly,
there holds the completeness relation
$\sum \limits_{{\sigma} \in \hat{G}} P^{\sigma} = 1$,
where the limit of infinite sums is understood in the strong
operator topology.

Next, we define maps $M^{\sigma}_{\iota}$ from $\Cal F(\Cal O)$ into
$\Cal F(\Cal O)$, setting
$$
M^{\sigma}_{\iota} (\psi) \doteq \int dg D^{\sigma}_{\iota\iota}
(g^{-1}) V(g) \psi V(g)^{-1}\ . \tag 4.3
$$
Since the matrices $D^{\sigma}_{\iota \kappa} (g^{-1})$ are unitary,
there
holds $|D^{\sigma}_{\iota \kappa} (g^{-1})| \leq 1$ and consequently
each map $M^{\sigma}_{\iota}$ is bounded in norm by 1. Hence the
linear map from $\Cal F(\Cal O)\times \Cal F(\Cal O)$ into itself,
which is given by
$$
\psi \times \psi{'} \longrightarrow M^{\sigma}_{\iota} (\psi) \times
M^{\sigma}_{\iota} (\psi'), \tag 4.4
$$
also has  norm less than or equal to 1. We compose the latter map
with $\Xi$ and thereby obtain a map $\Xi^{\sigma}_{\iota}$ from
$\Cal F(\Cal O) \times \Cal F(\Cal O)$ into $HS(\Cal H)$. This map
clearly has the same compactness properties as those stated for
$\Xi$ in Proposition 3.3.

Making use  of this fact we want to exhibit compactness properties
of the maps
$$
\Theta^{\sigma}_{\iota}\doteq P^{\sigma}_{\iota}\cdot\Theta\ .\tag
4.5
$$
In order to simplify the notation we keep $\sigma,\iota$ fixed for a
moment and put, for given $\psi \in \Cal F(\Cal O)$,
$$
\Psi \doteq P^{\sigma}_{\iota} e^{-\beta H}\psi \Omega\ .\tag 4.6
$$
The orthogonal projection onto the ray of $\Psi$ is denoted by
$E_{\Psi}$. It will be crucial in the following that the orbit of any
such vector $\Psi$ under the action of the gauge transformations
$V(g)$ gives rise to an irreducible representation of $G$.

With the above notation we can write
$$
\Xi^{\sigma}_{\iota} (\psi \times \psi^{\ast})
= \int dg \bigl( V(g) \Psi,\cdot\bigr) V(g) \Psi = \| \Psi
\|^2\cdot M(E_{\Psi})\ . \tag 4.7
$$
Hence, for any pair of operators $\psi,\psi'\in \Cal F(\Cal O)$, there
holds
$$
\bigl\| \Xi^{\sigma}_{\iota} (\psi \times \psi^{\ast} - \psi' \times
\psi{'}^{\ast})\bigr\|_{HS} = \bigl\| \| \Psi \|^2 M
(E_{\Psi}) - \| \Psi '\|
^2 M(E_{\Psi'})\bigr\|_{HS}\ . \tag 4.8
$$
An important lower bound for the right hand side of this inequality
is given in the following lemma.

\noindent\proclaim{Lemma 4.1} Let $\Psi,\Psi'\in
{\Cal H}^{\sigma}_{\iota}$.
Then
$$
\bigl\| \| \Psi \|^2 M(E_{\Psi})-\| \Psi' \|^2 M(E_{\Psi'})\bigr\|_{HS}
\geq {1\over \sqrt{2d_{\sigma}}} \inf_{\bar V} \| \Psi - \bar V \Psi' \|
^2\ ,
$$
where  the infimum is understood with respect to all unitary
operators
$$
\bar V \in \bigl\{ V(g) \upharpoonright {\Cal H}^{\sigma} : g \in
G\bigr \}''\ .
$$
\endproclaim\medskip

\noindent{\it Remark\/}: Since the restriction of the
representation $V$ of $G$ to the subspace ${\Cal H}^{\sigma} $ is
equivalent to a multiple of the irreducible representation
$\sigma$ it follows that the group generated by the unitaries $\bar
V$ in the statement is isomorphic to the group $\Cal U (d_{\sigma})$
of all unitaries on a $d_\sigma$--dimensional Hilbert space.\medskip

\noindent{\it Proof\/}: Since the operator $M(E_{\Psi})$ commutes
with $V(g),\,g\in G$, it follows from Schur's Lemma that it is a
multiple of the projection operator $I_{\Psi}$ which projects onto
the $d_{\sigma}$--dimensional irreducible subspace of the
representation $V$,
containing $\Psi$. Hence, by computing traces, one finds that
$M(E_{\Psi}) = d^{-1}_{\sigma}\cdot I_{\Psi}$. An analogous
statement holds for $\Psi'$. Bearing in mind the definition of the
scalar product in $HS(\Cal H)$ it follows that
$$
\align
\bigl\| \|  \Psi \| ^2 M &(E_{\Psi}) - \|  \Psi' \| ^2 M
(E_{\Psi'})\bigr\|
^2_{HS}=\\
&= d^{-1}_{\sigma}\| \Psi \| ^4 + d^{-1}_{\sigma} \| \Psi' \|^4 - 2
d^{-2}_{\sigma} \|  \Psi \| ^2 \|  \Psi' \| ^2 Tr I_{\Psi}  I_{\Psi'}\\
&\geq d^{-1}_{\sigma} \bigl ( \| \Psi \| ^4 +
\| \Psi' \|^4 - 2 \| \Psi \| ^2
\|  \Psi' \| ^2 \| I_{\Psi}  I_{\Psi'} \|^2\bigr) \\
&=  d^{-1}_{\sigma} \bigl ( \| \Psi \| ^4 + \| \Psi' \|^4 -
2 \sup_{\bar V}
|( \Psi, \bar V \Psi')|^2\bigr)\ .
\endalign
$$
Here we made use of the fact (in the inequality) that $|Tr I_{\Psi}
I_{\Psi'}|= Tr I_{\Psi} I_{\Psi'} I_{\Psi} \leq d_{\sigma} \cdot \|
I_{\Psi}
I_{\Psi'} \|^2$ and (in the last equality) that the unitaries $\bar V
\in \{V(g) \upharpoonright {\Cal H}^{\sigma} : g\in G\}''$ act
transitively on the unit ball of every irreducible subspace of
the representation $V$ in ${\Cal H}^\sigma$. By an elementary
calculation one sees that for
$Re(\Psi,\bar V \Psi') \geq 0$
$$
\align
{1\over 2} \| \Psi - \bar V \Psi' \|^4 &= {1\over 2} \bigl(\| \Psi \|^2
+ \| \Psi' \|^2 - 2 Re(\Psi,\bar V \Psi')\bigr)^2\\
&\leq \| \Psi \|^4 + \| \Psi' \|^4 - 2 \bigl( Re(\Psi, \bar V \Psi' )
\bigr)
^2\ .
\endalign
$$
Hence, since the set of unitaries $\bar V$ is stable under
multiplication with phase factors $\eta \in \Bbb C\ ,\ |\eta|=1$,
there holds
$$
{1\over 2} \inf_{\bar V} \| \Psi - \bar V \Psi' \|^2 \leq \| \Psi
\|^4 + \| \Psi'\|^4 - 2\sup_{\bar V} |(\Psi,\bar V \Psi')|^2\ .
$$
The statement then follows. \hfill $\qed$

Making use of this lemma and recalling that $\Psi =
P^{\sigma}_{\iota} e^{-\beta H} \psi \Omega = \Theta ^{\sigma}_{\iota}
(\psi)$, we can proceed from (4.8) to the estimate
$$
\sqrt{2d_{\sigma}}\  \| \Xi ^{\sigma}_{\iota} (\psi \times \psi ^{\ast} -
\psi' \times \psi{'}^{\ast})\|_{HS} \geq \inf_{\bar V} \| \Theta
^{\sigma}_{\iota} (\psi) - \bar V\cdot \Theta ^{\sigma}_{\iota}
(\psi')\|^2\tag 4.9
$$
which holds for any $\psi, \psi' \in \Cal F(\Cal O)$. We are now in
the position to establish the following result.

\noindent \proclaim{Proposition 4.2} Let Conditions C and J be
satisfied and let $\Theta ^{\sigma}_{\iota}$ be the map from $\Cal
F(\Cal O)$ into ${\Cal H}^{\sigma}_{\iota}$, defined in {\rm (4.5)}.
Then $\Theta
^{\sigma}_{\iota}$ is compact, and its $\varepsilon$--content is bounded
by
$$
N_{\Theta ^{\sigma}_{\iota}} (\varepsilon) \leq
\inf_{\varepsilon_1+\varepsilon_2={\varepsilon\over 2}}
\underline{N}
\biggl({\varepsilon ^2_1 \over \sqrt{2d_\sigma}}\biggr)\cdot
\biggl( {2d_{\sigma} e ^{-\beta m_{\sigma}}\over \varepsilon_2} +
1\biggr) ^{d^2_{\sigma}}, \varepsilon >0\ ,
 $$
where $m_{\sigma}$ is the lower boundary of the mass spectrum
in the sector $\sigma$ and $\underline{N} (\varepsilon)$ is defined
in Lemma 3.1.\endproclaim\bigskip

\noindent{\it Proof\/}: For the proof of the statement we make use
of the estimate (4.9) as well as of Lemma 2.6 and 2.7. Let us
assume that there are, for given $\varepsilon >0\, , \,
M(\varepsilon)$ elements $\psi_i \in \Cal F(\Cal O)_1\, ,\, i=1,\cdots
M(\varepsilon)$, such that $\inf\limits_{\bar V} \| \Theta
^{\sigma}_{\iota} (\psi_i)-{\bar V}\Theta ^{\sigma}_{\iota}(\psi_k)\|>
\varepsilon $ for $i\not= k$. It then follows from (4.9) that
$$
\| \Xi^{\sigma}_{\iota} \bigl(\psi_i\times \psi_i^{\ast} - \psi_k \times
\psi_k^{\ast}\bigr) \|_{HS} >  {\varepsilon^2\over
\sqrt{2d_{\sigma}}}\ ,\ i\not= k\ .
$$
But this implies, since $\psi_i\times \psi_i^{\ast} \in  \bigl(\Cal F(\Cal
O) \times \Cal F(\Cal O)\bigr)_1\ ,\ i = 1,\cdots M(\varepsilon)$,
that $M(\varepsilon)$ cannot be larger than the
$\varepsilon^2/\sqrt {2d_{\sigma}}$--content of
$\Xi^{\sigma}_{\iota}\ ,\ M(\varepsilon) \leq N_{\Xi
^{\sigma}_{\iota}} \biggl({\varepsilon^2\over
\sqrt{2d_{\sigma}}}\biggr)$. Next we recall that the set of unitaries
$\bar V \in  \bigl\{ V(g) \upharpoonright {\Cal H}^{\sigma} : g \in
G\bigr\}''$ forms a group $\bar {\Cal V}$ which is isomorphic to the
group of unitaries $\Cal U(d_{\sigma})$. Thus it follows from
Lemma 2.6 that the
$\varepsilon$--content of $\bar {\Cal V}$ is bounded by
$$
N^{\bar {\Cal V}} (\varepsilon) \leq \biggl( {2d_{\sigma}\over
\varepsilon} + 1\biggr)^{d^2_{\sigma}}\ ,\ \varepsilon>0\ .
$$
We also note that  $\| \Theta^{\sigma}_{\iota} (\psi)\| = \|
P^{\sigma}_{\iota}e^{-\beta H} \psi \Omega \| \leq e ^{- \beta
m_{\sigma}} \| \psi \|$ and consequently $\| \Theta
^{\sigma}_{\iota} \| \leq e ^{-\beta m _{\sigma}}$. We can now apply
Lemma 2.7 to the set $\Cal X = \Theta ^{\sigma}_{\iota} \bigl (\Cal F
(\Cal O )_1\bigr)$, giving
$$
N_{\Theta^{\sigma}_{\iota}} (\varepsilon) \leq
\inf_{\varepsilon_1+e^{-\beta m_{\sigma}} \varepsilon_2 =
\varepsilon/2} M(\varepsilon_1) \cdot N^{\bar \Cal V}
(\varepsilon_2)\ .
$$
Hence the statement follows from the preceding estimates,
Proposition 3.3, and the fact that the $\varepsilon$--content of
$\Xi^{\sigma}_{\iota}$ is not larger than that of $\Xi$. \hfill $\qed$

This result provides information on the compactness properties of
the components of the map $\Theta$ in the various superselection
sectors of the theory. Since, for $\psi \in \Cal F(\Cal O)$,
$$
\sum_{\sigma\in \hat G} \sum^{d_{\sigma}}_{\iota=1} \Theta
^{\sigma}_{\iota} (\psi) = \biggl (\sum_{\sigma \in \hat G} \sum
^{d_{\sigma}}_{\iota=1} P^{\sigma}_{\iota}\biggr ) \cdot \Theta (\psi)
= \Theta (\psi)\tag 4.10
$$
one obtains with the help of Lemma 2.3 a similar result for the
map $\Theta$ if there are only a finite number of superselection
sectors in the theory. In order to obtain information on $\Theta$
in the general case we make use of Condition P. As was discussed
in the proof of the preceding proposition, there holds $\| \Theta
^{\sigma}_{\iota} \| \leq e ^{-\beta m_{\sigma}}$, and consequently
$$
\sum_{\sigma\in \hat G} \sum^{d_{\sigma}}_{\iota=1}\|
\Theta^{\sigma}_{\iota}\| \leq \sum_{\sigma\in \hat G} d_{\sigma}
e ^{-\beta m_{\sigma}} = \Cal Z (\beta) < \infty \tag 4.11
$$
if Condition P is satisfied. Thus in this case $\Theta$ is equal to an
absolutely converging sum of compact maps and therefore it is
also compact. (As a matter of fact, this result follows from
considerably weaker versions of Condition P. It would suffice for
example if, for $\varepsilon > 0\ ,  e^{-\beta m_{\sigma}} <
\varepsilon$ for almost all superselection sectors $\sigma$.)

It is not quite as easy to estimate the $\varepsilon$--content of
$\Theta$. In order to abreviate the argument we do not aim here
at an optimal estimate and proceed as follows. We pick $\alpha > 0$
and consider the maps $e^{-\alpha H}\cdot \Theta
^{\sigma}_{\iota}$. Since $\| e^{-\alpha H} \cdot \Theta
^{\sigma}_{\iota} (\psi) \| \leq e^{-\alpha m_{\sigma}} \| \Theta
^{\sigma}_{\iota} (\psi)\|$ we obtain for the respective
$\varepsilon$--contents of these maps the inequality
$N_{e^{-\alpha H}\cdot \Theta ^{\sigma}_{\iota}} (\varepsilon) \leq
N_{\Theta ^{\sigma}_{\iota}} \bigl (e^{\alpha m _{\sigma}} \cdot
\varepsilon \bigr),\  \varepsilon > 0$. Next we consider, for fixed
$\sigma$,  the map $e^{-\alpha H} \cdot \Theta^{\sigma} \doteq
\sum \limits ^{d_{\sigma}}_{\iota=1} e^{-\alpha H}\cdot \Theta
^{\sigma}_{\iota}$. It follows  from Lemma 2.3, the preceding
bounds, and Proposition 4.2 that, for $\varepsilon >0$,
$$\align
&N_{e^{-\alpha H}\cdot \Theta ^{\sigma}}(\varepsilon) \leq
\prod^{d_{\sigma}}_{\iota=1} N_{\Theta ^{\sigma}_{\iota}}\left(
{e^{\alpha m _{\sigma}} \over
2d_{\sigma}} \varepsilon \right)\\
&\leq \inf_{\varepsilon_1+\varepsilon'_1={\varepsilon\over 2}}
\underline{N} \biggl ({e^{2\alpha m_{\sigma}} \over
(2d_{\sigma})^{5/2}} \varepsilon^2_1 \biggr)^{d_\sigma} \left (
{(2d_{\sigma})^2
e^{-(\alpha
+ \beta ) m_{\sigma}}\over \varepsilon'_1} + 1\right
)^{d^3_{\sigma}}\ .\tag 4.12
\endalign$$
Finally, we consider for finite subsets $\hat F \subset \hat G$
the maps $e^{-\alpha H}\cdot \Theta ^{\hat F} = \sum _{\sigma \in
\hat F} e^{-\alpha H} \cdot \Theta ^{\sigma}$. Since for $\psi \in \Cal
F (\Cal O)$
$$
e^{-\alpha H} \cdot \Theta ^{\hat F} (\psi) = \sum_{\sigma \in
\hat F} \sum^{d_{\sigma}}_{\iota=1} P^{\sigma}_{\iota}\cdot e^{-
(\alpha + \beta) H}\psi \Omega\tag 4.13
$$
it follows that $e^{-\alpha H}\cdot \Theta^{\hat F} (\psi)$
converges strongly to $e^{-\alpha H}\cdot \Theta (\psi)$ for any
increasing net $\hat F \nearrow \hat G$. Hence, for any such
net, the family of maps $e^{-\alpha H} \cdot \Theta ^{\hat F}$
dominates $e^{-\alpha H}\cdot \Theta$ asymptotically in the sense
of Lemma 2.4. So in order to get a bound for the
$\varepsilon$--content of $e^{-\alpha H}\cdot \Theta$ we have to
establish uniform bounds for the $\varepsilon$--contents of the
maps $e^{-\alpha H}\cdot \Theta^{\hat F}$. To this end we make
again use of Lemma 2.3, from which it follows that
$$
N_{e^{-\alpha H}\cdot \Theta^{\hat F}} (\varepsilon) \leq
\inf_{\varepsilon_1 + \cdots + \varepsilon_F = {\varepsilon\over
2}} N_{e^{-\alpha H}\cdot \Theta^{\sigma_1}} (\varepsilon_1) \cdots
N_{e^{-\alpha H}\cdot\Theta^{\sigma_F}} (\varepsilon_F)\ ,\tag 4.14
$$
where $F$ is the cardinality of $\hat F$ and $\sigma_i \in \hat F\,,\,
i=1, \cdots F$. Plugging into this estimate the preceding bounds on
$N_{e^{-\alpha H}\cdot \Theta^{\sigma}}$ we get
$$\align
N_{e^{-\alpha H}}&_{\cdot \Theta^{\hat F}} (\varepsilon)
\leq \inf_ {\varepsilon_1+ \cdots + \varepsilon_F+
\varepsilon'_1+ \dots + \varepsilon'_F= {\varepsilon\over 2}}
\prod^F_{i=1} \underline{N} \biggl({e^{2\alpha m_{\sigma_i}}
\over (2d_{\sigma_i})^{5/2} } \varepsilon^2_1
\biggr)^{d_{\sigma_i}}\times\\
&\times
\prod^F_{j=1}\left({(2d_{\sigma_j})^2e^{(\alpha+\beta)m_{\sigma_j}}\over
\varepsilon'_j} + 1\right) ^ {d^3_{\sigma_j}}\ .\tag 4.15
\endalign$$
We proceed to an upper bound of the right hand side of this
inequality by restricting  the infimum to the partitions
$\varepsilon_1+\cdots +\varepsilon_F=\varepsilon'_1+\cdots +
\varepsilon'_F={\varepsilon\over4}$. This allows us to treat
separately the ``kinematical factor'' in this expression.

\noindent\proclaim{Lemma 4.3} Let $0< r \leq 1$. There holds for
$\varepsilon >0$
$$
\inf_{\varepsilon'_1 +\cdots \varepsilon'_F= {\varepsilon \over 4}}
\prod ^F_{j=1}\left ( {(2d_{\sigma_j})^2e^{-(\alpha +
\beta)m_{\sigma_j}}\over \varepsilon'_j} + 1\right)
^{d^3_{\sigma_j}} \leq \exp\ \biggl ( {{1\over r} \Cal Z \left ( {r(\alpha +
\beta)\over 3+2r}\right) ^{3+2r}  \left ({16\over
\varepsilon}\right )^r \biggr )}
$$
uniformly for all finite subsets $\hat F\subset \hat
G$.\endproclaim\medskip

\noindent{\it Proof\/}: Since for fixed $r>0$ and any
$\lambda \geq 0$ there holds
$ln (1+\lambda) \leq r^{-1}\lambda ^r$ one gets for the left
hand side of the inequality in the statement the upper bound
$$
\inf_{\varepsilon'_1+\cdots + \varepsilon'_F = {\varepsilon\over 4}} \exp
\left ( {1\over r} \sum ^F_{j=1} d^3_{\sigma_j} {(2d_{\sigma_j})^{2r}
e^{-r(\alpha + \beta) m_{\sigma_j}}\over \varepsilon{'}^r_j}\right)\ .
$$
This infimum can be computed by an application of the Lagrange
multiplier method. It is equal to
$$
\exp \left( {1\over r} \biggl({16\over \varepsilon}\biggr )^r \biggl (
\sum ^F_{j=1} d_{\sigma_j}^{(3+2r)/(1+r)} e^{-r(\alpha +
 \beta)m_{\sigma_j}/ (1+r)}\biggr ) ^{(1+r)}\right )\ .
$$
Since $(3+2r)/(1+r)\geq 1$ the sum in this expression can be
estimated by $$\align
\biggl( \sum^F_{j=1} &d_{\sigma_j}^{(3+2r)/(1+r)}
e^{-r(\alpha+ \beta) m_{\sigma_j}/(1+r)}\biggr )^{(1+r)}\\
&\biggl( \sum^F_{j=1} d_{\sigma_j} e^{-r(\alpha+\beta)
m_{\sigma_j}/(3+2r)}\biggr )^{(3+2r)}\leq \Cal Z \biggl({r(\alpha
+\beta)\over (3+2r)}\biggr)^ {3+2r}\ .
\endalign
$$
Hence the statement follows.\hfill $\qed$

Thus the kinematical factor in our estimate of the
$\varepsilon$--content of $e^{-\alpha H}\cdot \Theta^{\hat F}$ will
in general develop in the limit $\hat F\nearrow \hat G$ an
essential singularity as $\varepsilon\searrow 0$. But it follows from our
result that this singularity is only of order 0 if Condition P is
satisfied.

Let us discuss now  the properties of the term involving the
factors $\underline{N}(\varepsilon)$. Since $\underline{N}(\varepsilon)$
is the {\it
limes inferior\/} of the $\varepsilon$--content of maps with norm
less than or
equal to 1, there holds $\underline{N}(\varepsilon) = 1$ if $\varepsilon
> 2$. It therefore follows from Condition P that
$$
\sup_{\hat F\subset \hat G}
\inf_{\varepsilon_1+\cdots+\varepsilon_F= {\varepsilon\over 4}}
\prod ^F_{i=1} \underline{N}\biggl({e^{2\alpha m_{\sigma_i}}
\over (2d_{\sigma_i})^{5/2} } \varepsilon^2_i \biggr)^{d_{\sigma_i}}
< \infty\ .\tag 4.16
$$
In order to verify this statement we note that the above infimum
increases if we proceed to the condition
$\varepsilon_1+\cdots+\varepsilon_F\leq \varepsilon/ 4$ since
$\underline{N}(\varepsilon)$ increases if $\varepsilon$ decreases.
Setting
$$
\varepsilon_i = \Cal Z \left( {2\over 5} \alpha \right)
^{-5/4} d_{\sigma_i}^{5/4} e^{-\alpha m_{\sigma_i}/2}\cdot
{\varepsilon \over 4},\ \quad i=1,\cdots F\tag 4.17
$$
there holds $(\varepsilon_1+\cdots+\varepsilon_F)\leq
{\varepsilon\over 4}$. Hence the left hand side of (4.16) is dominated by
$$
\sup_{\hat F\subset \hat G}\  \prod^F_{i=1} \underline{N} \biggl(
2^{-13/2} Z \biggl({2\over 5} \alpha\biggr)^{-5/2} e^{\alpha
m_{\sigma_i}}\varepsilon^2\biggr)^{d_{\sigma_i}}\ .\tag 4.18
$$
Now $e^{\alpha m_{\sigma_i}}$ is, for $\alpha > 0$, larger than
any given constant on almost all superselection sectors $\sigma$ if
Condition P is satisfied. Hence, for any $\varepsilon > 0$, only a
finite number of factors in the above products are different from
1, so the supremum exists. Consequently we get, for
$r>0$ and $\varepsilon>0$,
$$\align
N_{e^{- \alpha H}\cdot \Theta} (\varepsilon) \leq \text
{exp}\ \biggl({{1\over r}} &{\Cal
Z
\left( {r(\alpha + \beta)\over 3+2r}\right)^{3+2r}  \left({16\over
\varepsilon}\right)^r \biggr)}\times \\
&\times \prod_{\sigma \in \Sigma} \underline
N \left ( 2^{-13/2}\Cal Z \left( {2\over 5} \alpha\right)^{-5/2}
e^{\alpha m_{\sigma}}  \varepsilon^2\right)^{d_\sigma}\ .\tag 4.19
\endalign
$$
It remains to proceed from this bound for the
$\varepsilon$--content of the map $e^{-\alpha H} \cdot \Theta$ to
a corresponding bound for the map $\Theta$. This is accomplished
with the help of the following result whose proof is taken from
[3].

\noindent\proclaim{Lemma 4.4} Let $\alpha >0$. Then
$$
\| \Theta (\psi)\| \leq \| e^{-\alpha H}\cdot \Theta (\psi)\| ^{\beta/
(\alpha + \beta)}\cdot \| \psi \|^{\alpha/(\alpha+\beta)}, \quad
\psi \in
\Cal F (\Cal O)\ .
$$\endproclaim\medskip

\noindent{\it Proof\/}: Making use of the spectral theorem, we can
write
$$
\| \Theta (\psi)\|^2 = (\psi \Omega,e^{-2\beta H}\psi\Omega) =
\smallint e^{-2\beta\omega} d\mu(\omega)\ ,
$$
where $\mu$ is a measure with support on $\Bbb R_{\, +}$. Because of
H\"older's inequality and the fact that  $e^{-\lambda\omega}\leq 1$
for $\lambda,\omega \geq 0$ there holds
$$
\smallint e^{-2\beta \omega} d\mu(\omega) \leq\bigl( \smallint e
^{-2(\alpha + \beta) \omega} d\mu (\omega)\bigr)^{\beta/(\alpha
+\beta)} \bigl(\smallint d \mu (\omega)\bigr)^{\alpha/(\alpha +
\beta)}\ ,
$$
and since $\smallint d\mu(\omega) = \| \psi \Omega \|^2 \leq \| \psi
\|^2$, the statement follows.\hfill $\qed$\medskip

We can now state our main result.

\noindent\proclaim{Theorem 4.5} Let Conditions C, J and P be
satisfied. Then the maps $\Theta_{\beta,\Cal O}$, defined in (3.9),
are compact. Their $\varepsilon$--contents are, for any $0<\alpha
\leq \beta\ ,\ 0<r \leq 1$, bounded by
$$\align
N_{\Theta_{\beta,\Cal O}} (\varepsilon)&\leq
\text {exp}\ \left ({{2^5\over r}{\Cal Z}\left({r\over 5} \beta\right)^5
\left( {1\over \varepsilon} \right )^{r(1+\alpha/\beta)}}
\right )\times\\
&\qquad\qquad\times\prod_{\sigma \in \hat G} \underline{N} \left
(2^{-17/2} \Cal Z\biggl ({2\over 5}\alpha \biggr)^{-5/2}
e^{\alpha m_{\sigma}} \varepsilon^{2(1+\alpha/\beta)}\right)^{d_\sigma}\
, \quad \varepsilon > 0.
\endalign$$
Here $\Cal Z$ is the little partition function, defined in Condition P,
and $\underline{N}(\varepsilon)$ is the limes inferior of the
$\varepsilon$--contents of the maps $\Theta^{(0)}_{\beta,\Cal
O\cup(\Cal O+\underline x)}$, defined in Condition C, for
$\underline x$ tending to spacelike infinity.\endproclaim\medskip

\noindent{\it Remark\/}: The optimal choice of the parameters
$\alpha, r$ depends on the detailed properties of $\Cal Z$ and
$\underline{N}$.\medskip

\noindent{\it Proof\/}: According to Lemma 4.4 the
$\varepsilon$--content of $\Theta$ is not larger than the
$2^{- \alpha/\beta} \varepsilon^ {(1+\alpha/\beta)}$--content
of the map $e^{-\alpha H}\cdot \Theta$. Making use of the estimate
(4.19) on the $\varepsilon$--content of the latter map and the fact
that the little partition function $\Cal Z(\lambda)$ is monotonically
increasing if $\lambda$ decreases, the statement follows.\hfill $\qed$

The preceding theorem makes clear how the compactness
properties of the maps $\Theta^{(0)}_{\beta,\Cal O}$ determine the
compactness properties of their extensions $\Theta_{\beta,\Cal O}$ to
the local field algebras. It gives the desired information on the
general state of affairs, but in applications one is frequently
interested in a more explicit description of the compactness
properties of the respective maps. We therefore introduce the
following quantitative version of Condition C.\medskip

\noindent{\it Condition N$_p$\/}: Let $p>0\,,\,\beta>0$ be fixed
and let $\Cal O_1,\Cal O_2 \in \Cal K$. The maps $\Theta^{(0)}_{\beta,\Cal
O_1 \cup \Cal O_2}$, defined in Condition C, are
$p$--nuclear (for the given $p$) and
$$
\liminf_{\underline x} \| \Theta ^{(0)} _{\beta, \Cal O_1 \cup (\Cal
O_2 +\underline x)}\| _p < \infty
$$
for $\underline x$ tending to spacelike infinity.
\smallskip

Making use of this condition we can establish the following
interesting corollary.

\noindent\proclaim{Corollary 4.6} Let Conditions P, J and N$_p$ be
satisfied for some $0<p<{1\over 4}$. Then the maps
$\Theta_{\beta,\Cal O}$, defined in {\rm (3.9)}, are $q$--nuclear for
$q>{2p\over 1-2p}$. Moreover, if ${2p\over 1-2p}< q \leq 4p$
there holds
$$
\|\Theta_{\beta,\Cal O}\|_q \leq c_{p,q}\cdot \Cal Z \left( {1\over 15}
\biggl(q-{2p\over 1-2p} \biggr)\beta \right)^{5/2p} \cdot
\liminf_{\underline x} \| \Theta ^{(0)} _{\beta,\Cal O \cup (\Cal O +
\underline x)}\|^{1/2}_p\ ,
$$
where $c_{p,q}$ is some numerical constant.\endproclaim \medskip

\noindent{\it Proof\/}: We sketch the argument which is a
simple consequence of Proposition 2.5 and the preceding theorem.
One first notices that by Condition N$_p$, the third part of
Proposition 2.5 and relation (2.3) it follows that for $p'>p/(1-p)$
$$
\underline{N} (\varepsilon) = \liminf_{\underline x}
N_{\Theta^{(0)}_ {\beta, \Cal O \cup (\Cal O + \underline x)}}
(\varepsilon) \leq
e^{{\left ( const \cdot L/ \varepsilon \right )^{p'}}}\ ,
$$
where here and in the following $const$ stands for numerical
constants, depending only on the parameters $p,p'$, etc., and $L
\doteq \liminf\limits_{\underline x} \| \Theta^{(0)}_ {\beta, \Cal O \cup
(\Cal O + \underline x)}\|_p$. Plugging this information into the
statement
of the theorem and putting there $r=2p'$ one finds that for
$p/(1-p)<p'\leq 1/2$ and $0< \alpha \leq \beta$ there holds
$$
M_{\Theta_{\beta,\Cal O}} \bigl(2p' (1+\alpha/\beta)\bigr) \leq
const \  \Cal Z \left ({2\over 5} p\alpha\right)^{5/2p}\cdot
L^{1/2}\ .
$$
Here the monotonicity properties of $\Cal Z(\lambda)$ and the fact
that $\Cal Z(\lambda) \geq 1$ and $L\geq 1$ have been used in
order to simplify the expression. By applying to this estimate the
first two parts of Proposition 2.5 one arrives at
$$
\|\Theta_{\beta,\Cal O}\|_q \leq  const \  \Cal Z \left( {2\over
5} p\alpha \right)^{5/2p}\cdot L^{1/2}
$$
for $1\geq q > 2p
(1+\alpha/\beta)/\bigl(1-p(2+\alpha/\beta)\bigr)$. Since $\alpha$ can be
made arbitrarily small, the first part of the statement follows. The
quantitative estimate is obtained if one puts $\alpha/\beta =
{1\over 6p}\left(q -{2p\over 1-2p}\right) $. \hfill $\qed$\bigskip
\newpage

\noindent{\bf 5. Applications}

\noindent Starting from general, physically motivated assumptions we
have
established a tight relation between phase space properties of
observables and charged fields in relativistic quantum field
theory. These properties are encoded in specific features of the
maps $\Theta_{\beta,\Cal O \cup (\Cal O+\underline x)}^{(0)}$ which can be
described in qualitative and quantitative terms by compactness
or nuclearity conditions. We indicate here some applications of
these results to problems of physical interest and mention some
open questions.

One of the first applications of compactness respectively
nuclearity conditions has been the discussion in [1] of the
problem of causal (statistical) independence in relativistic quantum
field theory. On the mathematical side this problem amounts to
the question of whether the underlying net of observables or
fields has the so-called {\it split property\/} [18]. We recall that a
local net is said to have the split property if for each pair of
double cones $\Cal O_1$, $\Cal O_2$ such that the closure of $\Cal
O_1$ is contained in the interior of $\Cal O_2$ there exists a
factor $\Cal M$ of type $I_\infty$ (i.e., a von Neumann algebra which
is isomorphic to the algebra of all bounded operators on $\Cal H$)
such that $\gotO (\Cal O_1)\subset \Cal M \subset \gotO (\Cal
O_2)$, and similarly for the field algebras. It may happen that,
for given $\Cal O_1$, such factors $\Cal M$ exist only for
sufficiently large $\Cal O_2$. The net is then said to have the {\it
distal split property\/}.

It has been demonstrated in [1-5], cf. also [33], that the (distal)
split property of local nets is closely related to nuclearity
properties of the associated maps $\Theta_{\beta,\Cal O}^{(0)}$,
respectively $\Theta_{\beta,\Cal O}$. The present results show
that nuclearity properties of the maps $\Theta_{\beta, \Cal
O}^{(0)}$, involving the observables, imply that the field net has
the split property. We state this fact, which is a straightforward
consequence of Corollary 4.6, Proposition 2.5 and the results in
reference [5] in form of a theorem.

\proclaim{Theorem 5.1} Let Condition $P$, $J$ and $N_p$ be
satisfied for some $0<p<{1\over10}$. Then the field net $\Cal O \to
\Cal F(\Cal O)$ has the distal split property. If condition $N_p$ is
satisfied for all $p>0$ the field net has the split
property.\endproclaim
\smallskip

We note that this theorem has a partial converse. Namely, if the
field net has the distal  split property, then the maps
$\Theta_{\beta,\Cal O}$ are compact [3]. It is also noteworthy that
the split property of the field net is a direct consequence of the
split property of the observables if the gauge group is finite
abelian [18]. The split property of the field net has several
interesting consequences. We mention here only the fact that the
existence of local generators for internal, geometrical and supersymmetry
transformations can be established if this property holds [19,20].
One thereby
arrives at a rigorous version of ``current algebra''.

Another field of applications of compactness and nuclearity
conditions is the structural analysis of thermal states in
relativistic quantum field theory. In these applications one
frequently has to know how the nuclear $p$-norms of the
relevant maps depend on the size of the region $\Cal O$ and the
value of $\beta$.
According to the heuristic
considerations in [1,3], the $p$-norms may be interpreted as
partition functions. Hence, anticipating decent thermal properties,
it seems reasonable to expect (and can be established in models)
that
$$ \liminf_{\underline x}\Vert {\Theta_{\beta,\Cal O \cup
(\Cal O+\underline x)}^{(0)}}\Vert_p \le e^{cr^m\beta^{-n}}\tag5.1$$
for sufficiently large $r, \beta^{-1}$, where $r$ is the diameter of
$\Cal O $ and the (positive) constants $c$,$m$,$n$ may depend on
$p$. Similarly, there should hold for the little partition function
$$\Cal Z(\lambda) \leq e^{c \lambda^{-l}} \tag5.2$$
for small $\lambda >0$ and positive constants $c$ and $l$.
It then follows from Corollary 4.6 that a similar bound
as in (5.1) holds for the
norms
$\left\Vert\Theta _{\beta,\Cal O}\right\Vert _q, \, q>{2p\over 1-2p}$.
One can
therefore apply the results in [6] and establish the existence of thermal
equilibrium states for the smooth subnet \footnote{The
smooth subnet $\Cal F_0$ of $\Cal F$ is generated by all operators
$\psi\in \Cal F$
which transform norm continuously under time translations. It is
dense in $\Cal F$ in the strong operator topology.} $\Cal
F_0$ of the field net $\Cal F$. We recall that thermal equilibrium states are
distinguished by the fact that they satisfy the KMS-condition with respect to
time translations [34].

\proclaim {Theorem 5.2} Let Condition $J$ and the specific
versions {\rm (5.1)} and {\rm (5.2)} of Conditions $N_p$ and $P$ be
satisfied for
some
$0<p<{1\over4}$. Then there exist KMS-states for the (smooth)
field net $\Cal F_0$ for all positive temperatures
$\beta^{-1}>0$.\endproclaim
\medskip

Appealing to the thermodynamical interpretation of the nuclearity
conditions [1,3], this result can be rephrased in more physical
terms: if the partition function of the canonical ensemble with
zero total charge exists and exhibits the physically expected
behaviour (relation (5.1)) and if the mass
spectrum of the theory is sufficiently tame (relation (5.2)) then
the grand canonical ensemble with zero chemical potential exists
in the thermodynamical limit.

By a straightforward generalization of Condition $P$ this result
can be extended to the grand canonical ensembles with non-zero
chemical potential. To illustrate this fact let us assume that $Q$ is
the generator of a one-parameter subgroup of the gauge
transformations with eigenvalue $q_\sigma$ on the superselection
sectors $\sigma$  and let $\mu \in {\Bbb R}$ be such that the
function ${\Cal Z}_\mu ( \lambda ) = \sum_\sigma d_\sigma e^{-\lambda
(m_\sigma +\mu q_\sigma)}$ satisfies condition (5.2). (This would be
implied for sufficiently small $\mu$ by relation (5.2) if
the respective charge is tied to massive particles.) It then follows from
the
arguments given in Sec. 4 that the maps
$$ \Theta _{\beta,\mu,\Cal O}(\psi)=e^{-\beta (H+\mu
Q)}\psi\Omega, \quad \psi \in \Cal F (\Cal O)\ ,\tag5.3$$
are  $q$-nuclear and that conditions (5.1) and (5.2) lead to corresponding
upper bounds for their respective $q$-norms,
$q>{2p\over 1-2p}$.
Hence, by the results  in [6], there exist KMS-states for the
(smooth) field net and the dynamics
$U_\mu(t)=e^{it(H+\mu Q)}$, i.e., states with chemical potential
$\mu$ [35].

It is a remarkable fact that in theories of local observables the
existence of the grand canonical ensembles is implied  by the
existence of the (neutral) canonical ensemble and purely
kinematical conditions. This result originates from the fact that
the superselection
structure of the physical Hilbert space and the relation between fields and
observables is governed by a compact gauge group $G$ [24]. We
mention as an aside that the in a sense opposite problem of
whether one can derive from the partition functions relevant data
of the superselection sectors, such as their statistical dimensions
$d_\sigma$ (Kac-Wakimoto formulas [36]), has recently received
attention in the context of low dimensional theories, cf. for example
the discussion by B. Schroer [37]. Unfortunately, our estimates are too
weak to
make any general statements on this problem. But Proposition 4.2 seems to
support the idea that the statistical dimensions  $d_\sigma$ enter in a
universal way in the partition functions of the respective
canonical ensembles.

Let us mention in conclusion two interesting problems.

{(i)} It seems plausible that Conditions $P$ and $C$, respectively
$N_p$, are not completely independent. It would be of great
interest to {\it derive\/} information on the mass spectrum,
similar to Condition $P$, from suitable compactness or nuclearity
conditions involving the observables. A related question is: do
there exist theories where Conditions $C$ or $N_p$ are satisfied,
but the maps $\Theta_{\beta,\Cal O}$ of the field algebras into the
physical Hilbert space are not compact or nuclear?

{(ii)} Another interesting issue is the formulation of
compactness and nuclearity conditions in terms of the modular
operators of the theory, which are affiliated with the vacuum and the
local algebras [3]. This approach has the advantage that it can also
be applied to generally covariant quantum field theories where
time-translations are not a global spacetime symmetry and the
formulation of compactness conditions in terms of $H$ is no
longer possible. It would therefore be desirable to exhibit
physically significant conditions in terms of the modular
operators which allow one to derive nuclearity properties of
field algebras from corresponding properties of the observables.
A relevant step in this direction is the computation of modular
operators given in [38]. But one would need more specific
informations on the spectral properties of the relative modular
operators appearing in these formulas for the derivation of such
a result.
\vskip1.5cm

\noindent{\bf Acknowledgements.}
The authors thank S. Doplicher for discussions. D.B. is grateful for
the hospitality extended to him at various stages of this work
by the Dipartimento
di Matematica of the Universit\`a di Roma ``La Sapienza''
and ``Tor Vergata'' as
well as for financial support from Universit\`a di Roma and the
CNR.  C.D. would like to thank the members of the II. Institut f\"ur
Theoretische Physik, Universit\"at Hamburg for their hospitality.

\newpage
\baselineskip13pt

\noindent{\bf References}

\item {[1]} {\smc D. Buchholz, E.H. Wichmann}, ``Causal independence
and the energy--level density of states in quantum field theory'',
{\it Commun. Math. Phys.\/} {\bf 106}, 321--344 (1986).
\item {[2]} {\smc D. Buchholz, C.  D'Antoni, K. Fredenhagen},  ``The
universal structure of local algebras'', {\it Commun. Math. Phys.\/}
{\bf 111}, 123--135 (1987).
\item {[3]}  {\smc D. Buchholz, C. D'Antoni, R. Longo},``Nuclear maps
and modular structures II: Application to Quantum Field Theory'',
{\it Commun. Math. Phys.\/} {\bf 129}, 115--138 (1990).
\item {[4]} {\smc H.J. Borchers, R. Schumann},``A Nuclearity
condition for charged states'',{\it Lett. Math. Phys.\/} {\bf 23}, 63--77
(1991).
\item {[5]} {\smc D. Buchholz, J. Yngvason},``Generalized nuclearity
conditions and the split property in Quantum Field Theory'',  {\it
Lett. Math. Phys.\/} {\bf 23}, 159--167 (1991).
\item {[6]} {\smc D. Buchholz, P. Junglas},``On the existence of
equilibrium states in local quantum field theory'', {\it Commun.
Math. Phys.\/} {\bf 121}, 255--270 (1989).
\item {[7]} {\smc H. Narnhofer},``Entropy density for relativistic quantum
field
theory'' in {\it The state of matter\/}, eds. M. Aizenman, H. Araki,
Singapore,
World Scientific 1994.
\item {[8]} {\smc J. Bros, D. Buchholz},``Towards a relativistic KMS
condition'', {\it Nucl. Phys. B\/} (to appear).
\item {[9]} {\smc R. Haag, J. Swieca},``When does a quantum field
describe particles?'', {\it Commun. Math. Phys.\/} {\bf 1}, 308--320
(1965).
\item {[10]} {\smc V. En\ss},``Characterisation of particles by
means of local observables'', {\it Commun. Math. Phys.\/} {\bf 45},
35--52 (1975).
\item {[11]} {\smc D. Buchholz},``On particles,
infraparticles, and the problem of asymptotic completeness'' in {\it
VIIIth International Congress on Mathematical Physics\/},
Marseille 1986. Singapore, World Scientific 1987.
\item {[12]} {\smc D.Buchholz, M. Porrmann, U. Stein},  ``Dirac versus
Wigner: Towards a universal particle concept in local quantum
field theory'', {\it Phys. Lett.\/} {\bf B 267}, 377--381 (1991) .
\item {[13]}{\smc A. Einstein}, {\it The meaning of relativity\/}
(Appendix 2),
Princeton N.J. Princeton University Press 1955.
\item {[14]} {\smc R. Haag}, {\it Local Quantum Physics\/}, Berlin,
Heidelberg, New York, Springer 1992.
\item {[15]} {\smc D. Buchholz, P. Jacobi},  ``On the nuclearity
condition for massless fields'', {\it Lett. Math. Phys.\/} {\bf 13},
313--323 (1987).
\item {[16]} {\smc K. Fredenhagen, P. Hertel},
unpublished manuscript 1979.
\item {[17]} {\smc D. Buchholz, M. Porrmann},``How small is the
phase space in quantum field theory?'', {\it Ann. Inst. H.
Poincar\'e\/} {\bf 52}, 237--257 (1990) .
\item {[18]} {\smc S. Doplicher, R. Longo},``Standard and split
inclusions of von Neumann algebras'', {\it Invent. Mat.\/} {\bf 75},
493--536 (1984).
\item {[19]} {\smc S. Doplicher, R. Longo},``Local aspects of
superselection rules II'', {\it Commun. Math. Phys.\/} {\bf 88},
399--409 (1983).
\item{[20]} {\smc D. Buchholz, S. Doplicher, R. Longo},  ``On
Noether's theorem in quantum field theory'', {\it Ann. Phys.\/}
{\bf 170}, 1--17 (1986).
\item {[21]} {\smc S. Doplicher, R. Haag, J.E. Roberts},``Local
observables and particle statistics,'' I {\it Commun. Math. Phys.\/}
{\bf 23}, 199--230 (1971), II {\it Commun Math. Phys.\/} {\bf 35},
49--95 (1974).
\item {[22]} {\smc A. Pietsch}, {\it Nuclear locally convex spaces\/},
Berlin, Heidelberg, New York, Springer 1972.
\item {[23]} {\smc H. Jarchow}, {\it Locally convex spaces\/},
Stuttgart, Teubner 1981.
\item {[24]} {\smc S. Doplicher, J.E. Roberts},``Why there is a field
algebra with a compact gauge group describing the
superselection structure in particle physics'', {\it Commun. Math.
Phys.\/} {\bf 131}, 51--107 (1990).
\item {[25]} {\smc J.E. Roberts},  ``Spontaneously broken gauge
symmetries and superselection rules'' in  {\it Procedings of the
International School of Mathematical Physics\/}, Camerino 1974,
ed. G. Gallavotti, Universit\`a di Camerino 1976.
\item {[26]} {\smc D.Buchholz, S. Doplicher, R. Longo, J.E. Roberts},
``A new look at Goldstone's theorem'', {\it Rev. Math. Phys.\/} Special issue,
47--82 (1992).
\item {[27]} {\smc J.J. Bisognano, E.H. Wichmann},``On the duality
condition for a hermitean scalar field'', {\it J. Math. Phys.\/} {\bf
16}, 985--1007 (1975). ``On the duality condition for quantum
fields'' {\it J. Math. Phys.\/} {\bf 17}, 303--321 (1976).
\item {[28]} {\smc J. Yngvason},``A note on essential duality'', {\it
Lett. Math. Phys.\/} {\bf 31}, 127--141 (1994).
\item {[29]} {\smc D. Guido, R. Longo},``Relativistic
invariance and charge conjugation in quantum field theory'', {\it
Commun. Math. Phys.\/} {\bf 148}, 521--551 (1992).
 \item {[30]} {\smc D. Buchholz, S. Doplicher, R. Longo, J.E. Roberts},
``Extensions of automorphisms and gauge symmetries'', {\it
Commun. Math. Phys.\/} {\bf 155}, 123--134 (1993).
\item {[31]} {\smc D. Buchholz, K. Fredenhagen},``Locality and
structure of particle states'', {\it Commun. Math. Phys.\/} {\bf 84},
1--54 (1982).
\item {[32]} {\smc E. Hewitt, K. Ross}, {\it Abstract harmonic
analysis\/}
II, Berlin, Heidelberg, New York, Springer 1970 .
\item {[33]} {\smc C. D'Antoni, S. Doplicher, K. Fredenhagen, R.
Longo},``Convergence of local charges and continuity properties of
$W^{\ast}$--inclusions'', {\it Commun. Math. Phys.\/} {\bf 110},
325--348 (1987).
\item {[34]} {\smc O. Bratteli, D.W. Robinson}, {\it Operator algebras
and quantum statistical mechanics\/} II, Berlin, Heidelberg, New
York, Springer 1981.
\item {[35]} {\smc H. Araki, R. Haag, D. Kastler, M. Takesaki},
``Extension of KMS states and chemical potential'', {\it Commun.
Math. Phys.\/} {\bf 53}, 97--134 (1977)
\item {[36]} {\smc V.G. Kac, M. Wakimoto},``Modular and conformal
invariance constraints in representation theory of affine
algebras'', {\it Adv. Math.\/} {\bf 70}, 156--236 (1988).
\item{[37]} {\smc B. Schroer}, {\it Some useful properties of
rotational Gibbs states in chiral conformal quantum field theory\/},
preprint 1994 .
\item {[38]} {\smc T. Isola},  ``Modular structure of the crossed
product by a compact group dual'',
{\it J. Oper. Theory\/} (to appear).

\enddocument
\bye